\documentclass[11pt,a4paper]{article}
\usepackage{axodraw}
\usepackage{color}

\usepackage{epsfig,psfrag,graphicx}
\usepackage{amsfonts}
\usepackage{amssymb}
\usepackage{mathrsfs}
\usepackage{amsmath,amssymb}
\usepackage{subeqn}
\usepackage{bbm}
\usepackage[bookmarks=true]{hyperref}
\usepackage[nosort]{cite}
\usepackage[bulletsep]{collref}
\usepackage{verbatim}

\ifx\hypersetup\sadfkjashdfkxja\else
\hypersetup{pdftitle={Marginal Deformations and 3-Algebra Structures}}
\hypersetup{pdfauthor={Nikolas Akerblom, Christian Saemann, Martin Wolf}}
\hypersetup{pdfsubject={Mathematical Physics}}
\hypersetup{pdfkeywords={3-Algebras, Marginal Deformations, BLG, ABJM}}
\hypersetup{plainpages=false}
\hypersetup{bookmarksnumbered=true}
\hypersetup{pdfstartview=FitH}
\hypersetup{pdfpagemode=None}
\hypersetup{colorlinks=false}
\hypersetup{citebordercolor={.5 1 .5}}
\hypersetup{urlbordercolor={.5 1 1}}
\hypersetup{linkbordercolor={1 .7 .7}}
\fi

\makeatletter\renewcommand{\section}{\@startsection
{section}{1}{\z@}{-2.5ex plus -1ex minus
    -.2ex}{2.3ex plus .2ex}{\centering\large\bf\mathversion{bold}}}

\makeatletter\renewcommand{\subsection}{\@startsection{subsection}{2}{\z@}{-3.25ex
plus -1ex minus
   -.2ex}{1.5ex plus .2ex}{\centering\bf\mathversion{bold}}}

\makeatletter\renewcommand{\subsubsection}{\@startsection{subsubsection}{3}{-2.45ex}{-3.25ex
plus -1ex minus -.2ex}{1.5ex plus .2ex}{\centering\bf\mathversion{bold}}}

\makeatletter\renewcommand{\paragraph}{\@startsection{paragraph}{4}{\z@}%
                                    {0.8ex \@plus1ex \@minus.2ex}%
                                    {-.5em}%
                                    {\normalfont\normalsize\bfseries\mathversion{bold}}}

\renewcommand{\thesection}{\arabic{section}.}
\renewcommand{\thesubsection}{\arabic{section}.\arabic{subsection}.}

\numberwithin{paragraph}{section}
\renewcommand\theparagraph {\S\thesection\@arabic\c@paragraph.\kern-8pt}

\numberwithin{equation}{section}
\renewcommand{\theequation}{\thesection\arabic{equation}}

\renewcommand*\l@section{\@dottedtocline{1}{0em}{1.5em}}
\renewcommand*\l@subsubsection{\@dottedtocline{4}{3.8em}{3.2em}}

\renewcommand\tableofcontents{%
    \section*{\large\contentsname
        \@mkboth{%
          \MakeUppercase\contentsname}{\MakeUppercase\contentsname}}%
       {\baselineskip=15pt plus 2pt minus 1pt
    \@starttoc{toc}}%
}

\renewenvironment{thebibliography}[1]
     {\section*{\centering{\refname}
        \@mkboth{\MakeUppercase\refname}{\MakeUppercase\refname}}%
     \list{\@biblabel{\@arabic\c@enumiv}}%
           {\settowidth\labelwidth{\@biblabel{#1}}%
            \leftmargin\labelwidth
            \advance\leftmargin\labelsep
            \@openbib@code
            \usecounter{enumiv}%
            \let\p@enumiv\@empty
            \renewcommand\theenumiv{\@arabic\c@enumiv}}%
      \sloppy
      \clubpenalty4000
      \@clubpenalty \clubpenalty
      \widowpenalty4000%
      \sfcode`\.\@m
 \catcode`\^^M=10%
}

\newcommand{\appendices}{\section*{Appendices}\setcounter{subsection}{0}\setcounter{equation}{0}\renewcommand{\thesubsection}{\Alph{subsection}.}
\renewcommand{\theequation}{\thesubsection\arabic{equation}}
\addtocontents{toc}{\vspace{0.2cm}

{\bf Appendices}}
}

\def\slasha#1{\setbox0=\hbox{$#1$}#1\hskip-\wd0\hbox to\wd0{\hss\sl/\/\hss}}

\def\periodb#1{\setbox0=\hbox{$#1$}#1\hskip-\wd0\hbox to\wd0{-}}

\newcommand{\lbr}{(\hspace{-0.1cm}(}
\newcommand{\rbr}{)\hspace{-0.1cm})}
\newcommand{\lsb}{[\hspace{-0.05cm}[}
\newcommand{\rsb}{]\hspace{-0.05cm}]}
\newcommand{\CA}{\mathcal{A}}    			% cal-letters

\newcommand{\CCC}{\mathscr{C}}

\newcommand{\CL}{\mathcal{L}}

\newcommand{\CN}{\mathcal{N}}
\newcommand{\CO}{\mathcal{O}}

\newcommand{\CCO}{\mathscr{O}}

\newcommand{\CCP}{\mathscr{P}}

\newcommand{\CR}{\mathcal{R}}

\newcommand{\CW}{\mathcal{W}}

\newcommand{\frg}{\mathfrak{g}}				% frak-letters

\newcommand{\FR}{\mathbbm{R}}     			% field of real numbers
\newcommand{\FC}{\mathbbm{C}}     			% field of complex numbers
\newcommand{\RZ}{\mathbbm{Z}}     			% ring of integers
\newcommand{\dd}{\mathrm{d}}     			% total differential
\newcommand{\dpar}{\partial}     			% partial differential
\newcommand{\de}{\mathrm{e}}     			% Euler's number
\newcommand{\di}{\mathrm{i}}     			% imaginary unit
\newcommand{\eps}{{\varepsilon}}			% antisymmetric tensors
\newcommand{\Db}{{\bar{D}}}
\newcommand{\Phib}{{\bar{\Phi}}}

\newcommand{\bpsi}{{\bar{\psi}}}
\newcommand{\bth}{{\bar{\theta}}}
\newcommand{\bphi}{{\bar{\phi}}}
\newcommand{\bl}{{\bar{\lambda}}}

\newcommand{\bD}{{\bar{D}}}

\newcommand{\mdt}{{\dot{m}}}     			 \newcommand{\ndt}{{\dot{n}}}     			      			% dotted letters

\newcommand{\eand}{{~~~\mbox{and}~~~}}     		% and etc. in equations
\newcommand{\ewith}{{~~~\mbox{with}~~~}}
\newcommand{\efor}{{~~~\mbox{for}~~~}}

\newcommand{\tr}{\,\mathrm{tr}\,}     			% trace
\newcommand{\ao}{\mathfrak{o}}
\newcommand{\asu}{\mathfrak{su}}
\newcommand{\aso}{\mathfrak{so}}

\newcommand{\sU}{\mathsf{U}}     			% groups

\newcommand{\sSU}{\mathsf{SU}}
\newcommand{\sSL}{\mathsf{SL}}

\newcommand{\remark}[1]{}     				% remark
\def\tyng(#1){\hbox{\tiny$\yng(#1)$}}			% small Young diagram
\def\tyoung(#1){\hbox{\tiny$\young(#1)$}}			% small Young diagram
\begin{document}

\begin{titlepage}

\setcounter{page}{0}
\renewcommand{\thefootnote}{\fnsymbol{footnote}}

\begin{flushright}
NIKHEF/2009--008\\
TCDMATH 09--15\\
DAMTP 2009--47\\[.5cm]
\end{flushright}

\vspace*{1cm}

\begin{center}

{\LARGE\textbf{\mathversion{bold}Marginal Deformations and 3-Algebra Structures}\par}

\vspace*{1cm}

{\large
Nikolas Akerblom$^{a}$, Christian S\"amann$^{b}$
and Martin Wolf$^{c,}$\footnote{Also at the Wolfson College,
	Barton Road, Cambridge CB3 9BB, United Kingdom.}}\footnote{{\it E-mail
addresses:\/}
\href{mailto:nikolasa@nikhef.nl}{\ttfamily
nikolasa@nikhef.nl},~\href{mailto:saemann@maths.tcd.ie}{\ttfamily
saemann@maths.tcd.ie},~\href{mailto:m.wolf@damtp.cam.ac.uk}{\ttfamily
m.wolf@damtp.cam.ac.uk}}

\vspace*{1cm}

{\it $^{a}$ 
NIKHEF Theory Group\\
Science Park 105\\
1098 XG Amsterdam, The Netherlands}\\[.5cm]

{\it $^b$ Hamilton Mathematics Institute\\ \& \\
School of Mathematics\\
Trinity College, Dublin 2, Ireland}\\[.5cm]

{\it $^{c}$ Department of Applied Mathematics and Theoretical Physics\\
University of Cambridge\\
Wilberforce Road, Cambridge CB3 0WA, United Kingdom}

\vspace*{1cm}

{\bf Abstract}

\end{center}

\vspace*{-.3cm}

\begin{quote}

We study marginal deformations of superconformal
Chern-Simons matter theories that are based on 3-algebras. 
For this, we introduce the notion of an associated 3-product,
which captures very general gauge invariant deformations of
the superpotentials of the BLG and ABJM models. We also consider conformal multi-trace deformations preserving $\CN=2$
supersymmetry. 
We then use $\CN=2$ supergraph techniques to compute the two-loop
beta functions of these deformations. 
Besides confirming conformal invariance of both the BLG and ABJM models, we
also verify that the recently proposed $\beta$-deformations
of the ABJM model are indeed marginal to the order we are
considering.

\vfill
\noindent
June 09, 2009

\end{quote}

\setcounter{footnote}{0}\renewcommand{\thefootnote}{\arabic{thefootnote}}

\end{titlepage}

\tableofcontents

\bigskip
\bigskip
\hrule
\bigskip
\bigskip

\section{Introduction}

In the work of Bagger, Lambert \cite{Bagger:2006sk,Bagger:2007jr} as well as Gustavsson \cite{Gustavsson:2007vu}, a candidate theory for multiple M2-branes was proposed, which has attracted much attention in the last year. 
Initially, this theory was conjectured to be an IR description of stacks of M2-branes in the same sense as maximally supersymmetric Yang-Mills theory (SYM) provides an effective description of stacks of D-branes. Soon after its discovery, however, it was realized that this Bagger-Lambert-Gustavsson (BLG) model cannot capture stacks of arbitrarily many M2-branes: Its interactions and the gauge algebraic structure are based on 3-Lie algebras\footnote{See \cite{Lazaroiu:2009wz} and references therein for a detailed discussion of algebras with $n$-ary brackets.} \cite{Filippov:1985aa}, and there is only one such 3-Lie algebra which fulfills all reasonable physical requirements \cite{Nagy:2007aa,Papadopoulos:2008sk,Gauntlett:2008uf}.

One way to circumvent this problem is to generalize the concept of a 3-Lie algebra as done in \cite{Bagger:2008se} and \cite{Cherkis:2008qr}. These generalizations yield superconformal field theories which allow for more freedom but at the cost of a reduced amount of supersymmetry compared to the original BLG model. The generalizations discussed in \cite{Bagger:2008se}, for example, yield the so-called 
Aharony-Bergman-Jafferis-Maldacena (ABJM) model \cite{Aharony:2008ug} as a special case, 
see also \cite{VanRaamsdonk:2008ft}. This theory shares many features with 
$\CN=4$ SYM theory in four dimensions such as planar integrability
\cite{Minahan:2008hf,Gaiotto:2008cg,Gromov:2008qe,Bak:2008cp,Zwiebel:2009vb,Minahan:2009te,Bak:2009mq} 
(see \cite{Gaiotto:2007qi} for an earlier account). Therefore, it is interesting to ask what phenomena familiar from $\CN=4$ SYM theory in four dimensions persist in these generalized BLG-type models.

One such phenomenon is the existence of marginal deformations. There is a 3-parameter family of such deformations of $\CN=4$ SYM theory, which was found by Leigh and Strassler \cite{Leigh:1995ep}. These include in particular the so-called $\beta$-deformations as a subclass. Written in terms of four-dimensional $\CN=1$ superfields, where the field content of $\CN=4$ SYM theory is encoded in three chiral superfields $\Phi^i$, $i=1,2,3$, and a vector superfield, these deformations are given by the superpotential terms
\begin{equation}\label{eq:BetaDeformSYM}
 \CW\ =\ \eps_{ijk}\tr([\Phi^i,\Phi^j]_\beta \Phi^k)~,\ewith [\Phi^i,\Phi^j]_\beta\ :=\ \de^{\di\beta}\Phi^i\Phi^j-\de^{-\di\beta}\Phi^j\Phi^i~.
\end{equation}
The theories with such a superpotential are still finite and, as they are written in terms of superfields, they are manifestly $\CN=1$ supersymmetric.

In this paper, we make an attempt at the construction of analogous deformations for BLG-type models. For a rough guideline on what structures one expects to arise at 3-algebra level, one can look at the reduction process from M2-branes to D2-branes as described in \cite{Mukhi:2008ux}. For this reduction, one has to compactify a direction transverse to the M2-branes on a circle. In \cite{Mukhi:2008ux}, it was suggested that in this compactification process, the scalar describing M2-brane fluctuations in this direction would acquire the vacuum expectation value $\langle X^\circ\rangle=\frac{R}{\ell_p^{3/2}}=g_{YM}$.
Here, $R$ is the radius of the circle, $\ell_p$ the Planck length,  and $g_{YM}$ the Yang-Mills coupling constant. The interaction terms of the BLG model are formulated using totally antisymmetric 3-brackets of 3-Lie algebras. In the reduction process, 3-brackets of the form $[X^\circ,X^2,X^3]$ reduce to commutator terms $g_{YM}[X^2,X^3]=[X^\circ,X^2,X^3]$, and in a strong coupling expansion, only those 3-bracket expressions which reduce to a commutator survive. To obtain terms which correspond to $\beta$-deformed commutators, one evidently has to relax the total antisymmetry of the 3-bracket. One is therefore led to look for marginal deformations amongst models which are built from the 3-Lie algebras introduced in \cite{Bagger:2008se} and \cite{Cherkis:2008qr}. 

There are already proposals for $\beta$-deformations of both the BLG and the ABJM model in the literature \cite{Berman:2008be,Imeroni:2008cr} based on considering gravitational duals. Here, we will study such deformations in more detail from the gauge theory perspective: We will write down the most general gauge invariant deformations of BLG-type models based on 3-algebras in $\CN=2$ superspace. Although the generalized 3-Lie algebras of \cite{Bagger:2008se} and \cite{Cherkis:2008qr} already allow for certain classes of marginal deformations, we find that we should also introduce the notion of an associated 3-product: A new triple product, which transforms covariantly under gauge transformations. Moreover, we include all classically conformal multi-trace terms that are compatible with $\CN=2$ supersymmetry.\footnote{Multi-trace terms received attention in this context rather recently in \cite{Craps:2009qc}.}
The Lagrangians we find are rather restrictive, but contain the deformations studied in \cite{Imeroni:2008cr}. We then evaluate the beta functions of the couplings arising from the admissible deformations using supergraph techniques up to two-loop order. We confirm the conformal invariance of the BLG and the ABJM model as well as the deformations of \cite{Imeroni:2008cr} at quantum level to this order in perturbation theory.

This paper is structured as follows. In Section 2, we discuss the necessary 3-algebraic structures, the relation between 3-algebras and their associated gauge algebras and introduce associated 3-products. In Section 3, we present the Lagrangians of the BLG-type models we are interested in as well as their deformations. The results of our computation of the beta function up to two loops are then given in Section 4, and we conclude in Section 5.
In the Appendices, we collect some useful formul\ae{} used throughout this work.

\section{3-Algebras and associated 3-products}

The need for extending the BLG model to higher numbers of M2-branes led to two generalizations of the notion of a 3-algebra: the {\em generalized 3-Lie algebras} \cite{Cherkis:2008qr}, which we will refer to as {\em real 3-algebras}, and the {\em Hermitian 3-algebras} \cite{Bagger:2008se}, see also \cite{deMedeiros:2008zh} for a summary and a re-interpretation in terms of ordinary Lie algebras. In both cases, the underlying 3-bracket is no longer required to be totally antisymmetric. 

In the following, we will review these structures as well as their representations using matrix algebras. We also introduce the notion of an associated 3-product, a generalization of a 3-bracket\footnote{A similar generalization has been employed in \cite{Berman:2008be}.}, which will allow us to discuss extended superpotential terms yielding marginal deformations of both the BLG and ABJM models. 

\subsection{Real 3-algebras}

A {\em metric real 3-algebra} is a real vector space $\CA$ together with a trilinear bracket $[\cdot,\cdot,\cdot]\,:\,\CA\times\CA\times\CA\rightarrow \CA$ and a positive definite bilinear symmetric pairing $(\cdot,\cdot)\, :\, \CA\times\CA\rightarrow \FR$ satisfying the following properties for all $A,B,C,D,E\in\CA$:
\begin{subequations}
\begin{itemize}
\setlength{\itemsep}{-1mm}
 \item[(i)] The real fundamental identity:
\begin{equation}\label{Eq:FundamentalIdentity}
 [A,B,[C,D,E]]\ =\ [[A,B,C],D,E]+[C,[A,B,D],E]+[C,D,[A,B,E]]~,
\end{equation}
 \item[(ii)] the real compatibility relation:
\begin{equation}\label{Eq:Comp1}
 ([A,B,C],D)+(C,[A,B,D])\ =\ 0~,
\end{equation}
 \item[(iii)] and the real symmetry property:
\begin{equation}
(D,[A,B,C])\ =\ (B,[C,D,A])~.
\end{equation}
\end{itemize}
\end{subequations}
This is a generalization of the concept of a 3-Lie algebra in the sense of Filippov \cite{Filippov:1985aa}, which amounts to the special case of a totally antisymmetric 3-bracket. 

Choosing a basis $\tau_a$ of $\CA$, $a=1,\ldots,\mathrm{dim}\,\CA$, we can introduce the metric $h_{ab}$ and the structure constants $f_{abcd}$ as
\begin{equation}\label{eq:StructureConstantsReal}
h_{ab}\ :=\ (\tau_a,\tau_b)\eand f_{abcd}\ :=\ (\tau_d,[\tau_a,\tau_b,\tau_c])~. 
\end{equation}
Because of the properties (ii) and (iii), the structure constants obey the following symmetry relations:
\begin{equation}
 f_{abcd}\ =\ -f_{bacd}\ =\ f_{cdab}\ =\ -f_{abdc}~.
\end{equation}
When taking the 3-bracket of $\RZ_2$-graded objects as e.g.\ bosonic or fermionic fields, we define the 3-bracket to be insensitive to the grading:
\begin{equation}
 [A,B,C]\ :=\ A^aB^bC^c[\tau_a,\tau_b,\tau_c]~,\ewith A\ =\ A^a\tau_a~~\mbox{etc.}
\end{equation}

Every real 3-algebra comes with an associated Lie algebra $\frg_\CA$, 
the Lie algebra of {\em inner derivations} on $\CA$. Choosing a basis $\tau_a$ of $\CA$,
we define $\frg_\CA$ to be the image of the map $\delta\,:\,\Lambda^2\CA\rightarrow \mathrm{Der}(\CA)$ 
that is given by
\begin{equation}
\begin{aligned}
 &\Lambda^2\CA\ \ni\ X \ =\ X^{ab}\tau_a\wedge\tau_b\ \mapsto\ \delta_X\ \in\ \mathrm{Der}(\CA)\\
 &\kern1.7cm\delta_X(A)\ :=\ X^{ab}[\tau_a,\tau_b,A]
\end{aligned}
\end{equation}
for $A\in\CA$. Note that $X^{ab}=-X^{ba}$.
Note also that $\delta$ is not an injective map in general and thus the components $X^{ab}$ in the definition of $\delta_X$ are usually not uniquely defined. 
The Lie bracket $\lsb\cdot,\cdot\rsb$ on $\frg_{\mathcal{A}}$ is defined by the commutator action on $\CA$,
i.e.\ $\lsb\delta_X,\delta_Y\rsb(A):=\delta_X(\delta_Y(A))-\delta_Y(\delta_X(A))$ for $A\in\CA$. Closure of this bracket on $\frg_\CA$ follows from the fundamental identity \eqref{Eq:FundamentalIdentity}.

Additionally, we may endow the Lie algebra $\frg_\CA$ with a bilinear pairing 
\begin{equation}\label{eq:IPReal}
 \lbr \delta_X,\delta_Y\rbr\ :=\ X^{ab}Y^{cd}f_{abcd}~,
\end{equation}
which is symmetric, non-degenerate and $ad$-invariant, i.e.\ $\lbr\lsb\delta_X,\delta_Y\rsb,\delta_Z\rbr+
\lbr\delta_Y,\lsb\delta_X,\delta_Z\rsb\rbr=0$. 

The most prominent example of a 3-Lie algebra is the algebra $A_4$, which is the vector space $\FR^4$ endowed with the following 3-bracket and bilinear pairing:
\begin{equation}
 f_{abcd}\ =\ \eps_{abcd}\eand h_{ab}\ =\ \delta_{ab}~.
\end{equation}
The associated Lie algebra is $\frg_{A_4}\cong\aso(4)\cong\asu(2)\oplus\asu(2)$, and the bilinear pairing induced by the structure constants on this Lie algebra has split signature\footnote{This property is connected to parity invariance of the Chern-Simons Lagrangian, cf.\ Section 3.2.}: On the first $\asu(2)$ it is positive definite, on the second one negative definite. Further classes of examples of real 3-algebras are given in the next section.

\subsection{Matrix representations of real 3-algebras}

By a {\em matrix representation $\rho(\CA)$ of a 3-algebra $\CA$}, we will mean a homomorphism $\rho\,:\,\CA\rightarrow \CR:=\mathrm{Mat}(N,\FC)$, which forms a representation of the 3-algebra $\CA$ in the following way: The invariant pairing on $\CA$ is given by the natural scalar product $(A,B):=\tr(\rho(A)^\dagger \rho(B))$ for elements $A,B\in\CA$ and the 3-bracket is constructed using the natural operations on the matrix algebra: The product and the Hermitian conjugate. It should be stressed that $\rho(\CA)$ can be a true subset of $\CR$; however, the 3-bracket is certainly required to close on $\rho(\CA)$.

In the case of real 3-algebras, the matrix algebra $\CR$ is restricted\footnote{One could also choose Hermitian matrices; they, however, can be embedded into the real matrices, so that our restriction does not imply any loss of generality.} to $\mathrm{Mat}(N,\FR)$ and the Hermitian conjugate turns into the transpose. In the sequel, we will often not make a notational distinction between an element
$A\in\CA$ and its matrix realization $\rho(A)\in\CR$ and simply write $A$ in both cases.

Such representations have been classified in \cite{Cherkis:2008ha}, and for real 3-algebras, there are in fact four families:
\begin{equation}\label{3brackets}
 \begin{aligned}
\mathrm{I}^R_\alpha: &&& A,B,C\ \mapsto\ \alpha([[A^T,B],C]+[[A,B^T],C]+[[A,B],C^T]
   -[[A^T,B^T],C^T])~,\\
\mathrm{II}^R_{\alpha}: &&&
A,B,C\ \mapsto\ \alpha([[A,B^T],C]+[[A^T,B],C])~,\\
\mathrm{III}^R_{\alpha,\beta}: &&& A,B,C\ \mapsto\ \alpha(AB^T-BA^T)C+\beta C(A^T B-
 B^T A)~,\\
\mathrm{IV}^R_{\alpha,\beta}: &&& A,B,C\ \mapsto\ \alpha([[A,B],C]+[[A^T,B^T],C]+
 [[A^T,B],C^T]+[[A,B^T],C^T])\\
&&&~~~~~~~~~~~~~~+\beta([[A,B],C^T]+[[A^T,B],C]+[[A,B^T],C]+[[A^T,B^T],C^T])~,
 \end{aligned}
\end{equation}
where $\alpha$ and $\beta$ are arbitrary (real) parameters. Although $\alpha$ can always be removed from the bracket by a rescaling, we will find it convenient to keep it explicitly.

Besides forming representations, these brackets give rise to a real 3-algebra structure on $\mathrm{Mat}(N,\FR)$, and we denote the arising real 3-algebras by $M^R_{\rm I_\alpha}(N),\ldots,M^R_{\rm IV_{\alpha,\beta}}(N)$.

The case $M^R_{\rm III_{\alpha,\beta}}(N)$ is of particular importance: The real
3-algebras $\CCC^{2d}$ defined in \cite{Cherkis:2008qr} allow for
representations in the class ${\rm III}^R_{\alpha,\beta}$. The 3-Lie algebra
$A_4$, which is a sub-3-algebra of $\CCC^{4}$ can be identified
with a real sub-3-algebra of $M^R_{\rm III_{1,-1}}(4)$. Let us therefore expose
the associated Lie algebra structure of $M^R_{\rm III_{\alpha,\beta}}(N)$ in the
following. A derivation $\delta_X\in\frg_\CA$ acts on an element
$C\in\CA=M^R_{\rm III_{\alpha,\beta}}(N)$ according to
\begin{equation}
\begin{aligned}
 \delta_X(C)\ &=\ X^{ab}[\tau_a,\tau_b,C]\ =\ \underbrace{\alpha X^{ab}(\tau_a\tau_b^T-\tau_b\tau_a^T)}_{\ =:\ \hat X_L}C+C\underbrace{\beta X^{ab}(\tau_a^T\tau_b-\tau_b^T\tau_a)}_{\ =:\ \hat X_R}\\
&=\ \hat{X}_L C+C\hat{X}_R~.
\end{aligned}
\end{equation}
Thus, $\frg_\CA$ splits into two parts: one acting on $\CA$ from
the left and one acting from the right.  
The fact that $\frg_\CA$ forms a Lie algebra follows from the fundamental
identity as mentioned above. In particular,
\begin{equation}
 \lsb\delta_X,\delta_Y\rsb(C)\ =\ [\hat{X}_L,\hat{Y}_L]
C+C[\hat{Y}_R,\hat{X}_R]\ =\ \hat{Z}_L
C+C\hat{Z}_R\ =\ \delta_Z(C)~.
\end{equation}
Note that $\hat{X}_L=-\hat{X}_L^T$ and $\hat{X}_R=-\hat{X}_R^T$, that is, both
are antisymmetric matrices and they can be chosen independently. We therefore
conclude that $\frg_\CA\subseteq\ao(N)\oplus\ao(N)$ and in particular, if
$\rho(\CA)=\CR$, we have $\frg_\CA\cong\ao(N)\oplus\ao(N)$. 
Moreover, a short calculation reveals that the pairing on $\frg_\CA$ is given by
\begin{equation}\label{signature-real}
 \lbr X,Y \rbr\ =\ 
 X^{ab}Y^{cd}f_{abcd}\ =\ -\alpha\tr(\hat{X}^\dagger_L\hat{Y}_L)-\beta\tr(\hat{X}^\dagger_R\hat{Y}_R)~,
\end{equation}
and thus for $\alpha=-\beta$, the pairing has split signature. This property is required to render a Chern-Simons matter theory based on this gauge algebra parity invariant, see Section \ref{sec:deform}

\subsection{Associated 3-products of real 3-algebras}\label{sec:Assprodreal}

In gauge theories, the gauge potential (and its superpartners) takes values in a Lie algebra, while the matter fields take values in a representation of this Lie algebra. If the matter fields $X,Y$ sit in the adjoint matrix representation, there is a product between these fields -- the ordinary matrix product -- which transforms covariantly under gauge transformations $\delta_\Lambda=[\Lambda,\cdot]$:
\begin{equation}
 [\Lambda,X\cdot Y]\ =\ [\Lambda,X]\cdot Y+ X\cdot [\Lambda,Y]~.
\end{equation}
Both the matrix product and the commutator are special cases of the more general product
\begin{equation}
 \alpha_1 XY-\alpha_2YX~,\ewith\alpha_{1,2}\ \in\ \FC~,
\end{equation}
which also transforms covariantly. An analogous product can be introduced for representations of 3-algebras: Consider a matrix representation $\CR$ of a real 3-algebra $\CA$. An {\em associated 3-product} of $\CA$ in $\CR$ is a trilinear map $\langle A,B,C \rangle\,:\,\CR\times\CR\times\CR\rightarrow \CR$ satisfying the following identity:
\begin{equation}
 [A,B,\langle C,D,E\rangle]\ =\ \langle [A,B,C],D,E\rangle+\langle C,[A,B,D],E\rangle+\langle C,D,[A,B,E]\rangle~.
\end{equation}
This identity corresponds to the condition that the associated 3-product transforms covariantly under gauge transformations governed by the 3-bracket. Later on, this will allow us to replace ordinary 3-brackets in the superpotential by associated 3-products preserving gauge invariance. Evidently, all matrix representations of 3-brackets satisfy this identity and thus they are just special cases of associated 3-products. The general associated 3-product, however, allows for more general deformations of the superpotential than the conventional 3-bracket would do. In the Hermitian case, this includes in particular the deformations studied in \cite{Imeroni:2008cr}, as discussed later.

One may now ask for the most general 3-product, which can be written down using nothing but matrix products and transpositions, analogously to the matrix representations of 3-brackets \eqref{3brackets}. In the representation $\CR$ of type ${\rm III}_{\alpha,\beta}^R$, the most general such product reads as
\begin{equation}
 \langle A,B,C\rangle\ =\
\alpha_1 A B^T C+ \alpha_2CB^TA+\beta_1 BC^TA+\beta_2 AC^TB+\gamma_1 CA^TB+\gamma_2 BA^TC~,
\end{equation}
where $\alpha_{1,2}$, $\beta_{1,2}$ and $\gamma_{1,2}$ are real parameters.

\subsection{Hermitian 3-algebras}

A {\em metric Hermitian 3-algebra} is a complex vector space $\CA$ together with a bilinear-antilinear triple product $[\cdot,\cdot\,;\cdot]\,:\,\CA\times\CA\times\CA\rightarrow \CA$ and a positive definite Hermitian pairing\footnote{We choose the first slot to be antilinear and the second one to be linear.} $(\cdot,\cdot)\,:\,\CA\times\CA\rightarrow \FC$ satisfying the following properties for all $A,B,C,D,E\in\CA$:
\begin{subequations}\label{eq:AxiomsH3A}
\begin{itemize}
\setlength{\itemsep}{-1mm}
 \item[(i)] The Hermitian fundamental identity:
\begin{equation}\label{Eq:HFundamentalIdentity}
 [[C,D;E],A;B]\ =\ [[C,A;B],D;E]+[C,[D,A;B];E]-[C,D;[E,B;A]]~,
\end{equation}
 \item[(ii)] the Hermitian compatibility relation:
\begin{equation}\label{Eq:HComp1}
 (D,[A,B;C])-([D,C;B],A)\ =\ 0~,
\end{equation}
 \item[(iii)] and the Hermitian symmetry property:
\begin{equation}\label{sym}
(D,[A,B;C])\ =\ -(D,[B,A;C])~.
\end{equation}
\end{itemize}
\end{subequations}
With respect to a basis $\tau_a$ of $\CA$, we introduce the metric and the structure constants 
\begin{equation}
 h_{ab}=(\tau_a,\tau_b)\eand f_{abcd}\ :=\ (\tau_d,[\tau_a,\tau_b;\tau_c])~,
\end{equation}
which satisfy the following symmetry relations:
\begin{equation}
h_{ab}=(h_{ba})^*\eand
 f_{abcd}\ =\ -f_{bacd}\ =\ -f_{abdc}\ =\ (f_{cdab})^*~.
\end{equation}

Analogously to the case of real 3-algebras, a Hermitian 3-algebra comes with an associated Lie algebra, which is naturally a complex Lie algebra $\frg_\CA^\FC$. Here, we will merely be interested in a real form $\frg_\CA$ of $\frg_\CA^\FC$ that is defined as follows: Consider a basis $\tau_a$ of $\CA$ together with a basis $\tau_a^*$ of the complex conjugate $\CA^*$ of $\CA$.\footnote{The precise definition of $\CA^*$ is irrelevant at this point.} An element  $X=X^{ab}\tau_a\wedge\tau_b^*$ of 
$\mathfrak{Re}(\CA\wedge\CA^*)$ has components $X^{ab}$ satisfying $X^{ab}=-(X^{ba})^*$, and we 
then define $\frg_\CA$ to be the image of the map  $\delta\,:\,\mathfrak{Re}(\CA\wedge\CA^*) \rightarrow \mathrm{Der}(\CA)$, with $X\mapsto \delta_X$ and 
\begin{equation}
 \delta_X(A)\ :=\ X^{ab}[A,\tau_a;\tau_b]~,
\end{equation}
for $A\in\CA$. The Lie bracket $\lsb\cdot,\cdot\rsb$ on 
$\frg_\CA$ is defined as the commutator action of two inner derivations $\delta_X,\delta_Y\in\frg_\CA$ on $A\in\CA$. As in the case of real 3-algebras, closure of this bracket on $\frg_\CA$ follows from the fundamental identity.

A pairing on $\frg_\CA$ can be chosen as\footnote{Note that our definition differs from that of \cite{deMedeiros:2008zh} in that we have introduced an additional factor of $1/2$.} \cite{deMedeiros:2008zh}
\begin{equation}
\begin{aligned}
 \lbr \delta_X,\delta_Y\rbr\ &:=\ X^{ab}Y^{cd}\,f_{cabd}~.
\end{aligned} 
\end{equation}
This pairing is symmetric, bilinear, non-degenerate and $ad$-invariant. Note that when $\CA$ is considered as the carrier space for a representation of $\frg_\CA$, $\CA^*$ forms the carrier space for the complex conjugate representation.

\subsection{Matrix representations of Hermitian 3-algebras}

Let us now come to matrix representations of Hermitian 3-algebras as introduced in Section \ref{sec:Assprodreal} It was shown in \cite{Cherkis:2008ha} that there is only one such family of representations given by a
homomorphism $\rho\,:\,\CA\rightarrow {\rm Mat}(N,\FC)$ and the 3-bracket
\begin{equation}\label{Herm_I}
\mathrm{I}^H_\alpha:~~~ A,B,C\ \mapsto\ \alpha(AC^\dagger B-BC^\dagger A)~,
\end{equation}
where $\alpha$ is a real parameter. Interestingly, this is also the representation used in \cite{Bagger:2008se} to recast the ABJM model in 3-algebra language. 

In the following, we will denote the Hermitian 3-algebra defined by the above bracket on $\mathrm{Mat}(N,\FC)$ by $M_{\mathrm{I}_\alpha}^H(N)$. Note that the 3-Lie algebra $A_4$ introduced above coincides with the Hermitian 3-algebra $M_{\mathrm{I}_\alpha}^H(2)$.

The associated Lie algebra structure of this Hermitian 3-algebra is easily found to be $\frg_{\CA}\cong\asu(N)\oplus \asu(N)$, cf.\ \cite{Bagger:2008se}: Consider an element of $\delta_X=X^{ab}[\cdot,\tau_a;\tau_b]\in \frg_\CA$, where $\tau_a$ and $\tau_b$ are complex $N\times N$-matrices and $(X^{ab})^*=-X^{ba}$. With the definition \eqref{Herm_I}, we obtain ($\alpha=1$)
\begin{equation}
 \delta_X(A)\ =\ X^{ab}[A,\tau_a;\tau_b]\ =\ X^{ab}(A\tau_b^\dagger\tau_a-\tau_a\tau_b^\dagger A)~.
\end{equation}
Analogously to the case of $M^R_{\rm III_{\alpha,\beta}}(N)$, we can associate the following matrices with the inner derivations:
\begin{equation}
 \hat{X}_R\ =\ X^{ab}\tau_b^\dagger \tau_a\eand\hat{X}_L\ =\ -X^{ab}\tau_a \tau_b^\dagger~,
\end{equation}
which are both anti-Hermitian; for example, we have $(\hat{X}_R)^\dagger=(X^{ab}\tau_b^\dagger\tau_a)^\dagger=-X^{ba}\tau_a^\dagger\tau_b=-\hat{X}_R$. Similar considerations as in the real case show that $\hat{X}_R$ and $\hat{X}_L$ can be chosen independently, exhausting the fundamental representation of $\asu(N)$. The trace part is excluded as it would have a trivial action on $\CA$. Since left- and right-actions commute, we arrive at the conclusion that $\frg_\CA\cong\asu(N)\oplus \asu(N)$.

The symmetric bilinear pairing of elements $\delta_X,\delta_Y\in\frg_\CA$ is then given by
\begin{equation}
 \lbr X,Y\rbr\ =\ X^{ab}Y^{cd} f_{cabd}\ =\ 
 \tr(\hat{X}_L^\dagger\hat{Y}_L)-\tr(\hat{X}_R^\dagger\hat{Y}_R)~,
\end{equation}
and this expression shows that the signature on $\frg_\CA$ is again split, with positive and negative signature on the left and right acting subalgebra of $\frg_\CA$, respectively.

\subsection{Associated 3-products for Hermitian 3-algebras}

Consider again a matrix representation $\CR$ of a Hermitian 3-algebra $\CA$. By an associated 3-product of $\CA$ in $\CR$, we mean a bilinear-antilinear map $\langle A,B;C \rangle\,:\,\CR\times\CR\times\CR\rightarrow \CR$ satisfying the following identity:
\begin{equation}
 [\langle C,D;E\rangle,A;B]\ =\ \langle[C,A;B],D;E\rangle+\langle C,[D,A;B];E\rangle-\langle C,D;[E,B;A]\rangle~.
\end{equation}
We specialize now to the Hermitian 3-algebra $M_{\rm I_\alpha}^H(N)$ with basis $\tau_a$ for which $\CR={\rm Mat}(\FC,N)$. Note that the $\tau_a$ form a basis for both $M_{\rm I_\alpha}^H(N)$ and $\CR$. With respect to this basis, we can introduce structure constants of the associated 3-product as follows:
\begin{equation}
 \langle \tau_a,\tau_b;\tau_c\rangle\ =\ g_{abc}{}^d \tau_d\eand g_{abcd}\ =\ g_{abc}{}^e h_{de}~.
\end{equation}
In the representation $\CR$ of type $\mathrm{I}^H_\alpha$, the most general such product written in terms of matrices and Hermitian conjugation is given by the following expression:
\begin{equation}
 \langle A,B;C\rangle\ =\ \alpha_1 A C^\dagger B-\alpha_2 B C^\dagger A~,
\end{equation}
where $\alpha_{1,2}$ are complex parameters.

Below, we shall solely be interested in the one-parameter family that is given by $\alpha_1=\de^{\di\beta}$ and 
 $\alpha_2=\de^{-\di\beta}$ for $\beta\in\FR$. In analogy to the $\beta$-deformed commutator
given in \eqref{eq:BetaDeformSYM}, we denote the {\it $\beta$-3-bracket} by
\begin{equation}\label{eq:BetaCommutator}
  [\tau_a,\tau_b;\tau_c]_\beta\ :=\
\de^{\di\beta}\tau_a \tau_c^\dagger \tau_b-\de^{-\di\beta} \tau_b \tau_c^\dagger \tau_a 
  \ =:\ \big[\cos\beta {f_{abc}}^d+\di\sin\beta\, {d_{abc}}^d\big]\tau_d~.
\end{equation}
The ${f_{abc}}^d$ are the structure constants of the Hermitian 3-bracket and 
$d_{abcd}={d_{abc}}^e h_{de}$ obeys
\begin{equation}
 d_{abcd}\ =\ d_{bacd}\ =\ d_{abdc}\ =\ (d_{cdab})^*~.
\end{equation}
Therefore, 
\begin{equation}
 g_{abcd}\ =\ g_{badc}\ =\ -(g_{dcab})^*~.
\end{equation}
These symmetry properties of the structure constants $g_{abcd}$ can be re-phrased without 
referring to a particular choice of basis analogously to \eqref{Eq:HComp1} and \eqref{sym}:
\begin{equation}\label{eq:BetaDeformSCS}
\begin{aligned}
 (D,[A,B;C]_\beta)\ =\ -([D,C;A]_\beta,B)\eand
(D,[A,B;C]_\beta)\ =\ (C,[B,A;D]_\beta)~.
\end{aligned}
\end{equation}

Interestingly, \eqref{eq:BetaCommutator} will yield precisely the marginal deformations of the ABJM case 
recently studied in \cite{Imeroni:2008cr}.

\section{Deformations of BLG-type actions preserving $\CN=2$ supersymmetry}

In the following, we present deformations of BLG-type actions which make use of either real 3-algebras or Hermitian 3-algebras as their gauge 3-algebra structures. We will refer to these two cases as the real and Hermitian cases, respectively. All deformations will be manifestly $\CN=2$ supersymmetric and
supergauge invariant.

\subsection{Conventions}

We shall use the usual superfield conventions of \cite{Wess:1992cp} dimensionally reduced from four to three dimensions as done in \cite{Cherkis:2008qr}. Our superfields will live on $\FR^{1,2|4}$ and their expansions are given by
\begin{subequations}\label{eq:WZGaugeComponents}
\begin{equation}
 \Phi^i(y)\ =\ \phi^i(y)+\sqrt{2} \theta \psi^i(y)+\theta^2 F^i(y)~,
\end{equation}
for the chiral superfield and
\begin{equation}
 V(x)\ =\ - 
\theta^\alpha\bth^\beta(\sigma^\mu_{\alpha\beta}A_\mu(x)+\di\eps_{\alpha\beta}\sigma(x))+\di\theta^2(\bth\bar{\lambda}(x))-\di\bth^2(\theta\lambda(x))+\tfrac{1}{2}\theta^2\bth^2 D(x)
\end{equation}
\end{subequations}
for the vector superfield in Wess-Zumino (WZ) gauge.\footnote{When discussing the quantum theory, we will
not fix WZ gauge; see below.}
Here, $y$ are chiral coordinates, $i,j,\ldots=1,\ldots,N_f$ are flavor indices (counting complex field components) and $\alpha,\beta,\ldots=1,2$ are three-dimensional spinor indices. We will mostly be interested in $N_f=4$, but keeping $N_f$ arbitrary will prove useful as a book-keeping device. Notice that the spin group in $1+2$ dimensions is $\sSL(2,\FR)$ and hence, we do not need to distinguish between dotted and undotted spinors. In particular, indices of barred spinors can be contracted with those of unbarred ones. Our conventions for spinor contractions are as follows: $\chi\psi:=\chi^\alpha\psi_\alpha$,
$\bar\chi\bar\psi:=\bar\chi_\alpha\bar\psi^\alpha$. Furthermore, $\sigma^\mu$ are the $\sigma$-matrices in three dimensions
with $\sigma^\mu_{\alpha\beta}=\sigma^\mu_{\beta\alpha}$
and $\varepsilon_{\alpha\beta}=-\varepsilon_{\beta\alpha}$ with $\varepsilon_{\alpha\gamma}\varepsilon^{\gamma\beta}=\delta_\alpha^\beta$.

The superfields $\Phi^i$ take values in a 3-algebra\footnote{i.e.\ either a real or a Hermitian 3-algebra} $\CA$, while $V$ takes values in its associated Lie algebra $\frg_\mathcal{A}$. By a bar, we shall mean the appropriate complex conjugation operation (i.e.\ that of components and that of the gauge algebra representation).

To make our notation more concise, we shall always write $X(A)$ or even $X A$ as a shorthand for the action of an element $\delta_X$ of the associated Lie algebra $\frg_\CA$ on $A\in\CA$.

\subsection{Deformations of the superfield action in the real case}\label{sec:deform}

We start from a Wess-Zumino model minimally coupled to a Chern-Simons theory. Correspondingly, the superfield action reads as 
\begin{equation}\label{eq:S0Real}
 S^R_0\ =\ \di\sqrt{\kappa}\int \dd^{3|4}z\int_0^1\dd t\,\lbr V,\bar{D}^\alpha \big(\de^{-\frac{2\di}{\sqrt{\kappa}}tV}
 D_\alpha \de^{\frac{2\di}{\sqrt{\kappa}}tV}\big)\rbr+
 \int \dd^{3|4}z\,(\bar{\Phi}_i,\de^{-\frac{2\di}{\sqrt{\kappa}}V}\Phi^i)~,
\end{equation}
where $\dd^{3|4}z:=\dd^3x\,\dd^4\theta$, cf.\ \cite{Zupnik:1988en,Ivanov:1991fn,Cherkis:2008qr}.
The superfields $\Phi^i$ are all in the same representation of the gauge algebra $\frg_\CA$ whose carrier space is $\CA$. The coupling constant $\kappa$ is related to the Chern-Simons level $k$ via
$\kappa=k/\pi$. Notice that the vector superfield has been rescaled
appropriately to ensure that the action \eqref{eq:S0Real} has a proper
free-field limit,  $1/\sqrt{\kappa}\to0$, needed for perturbation theory. 

Recall that the ordinary Chern-Simons Lagrangian containing the Killing form of the Lie algebra as bilinear pairing receives a total sign under parity transformations. Many real 3-algebras, however, come with an associated Lie algebra of the form $\frg_\CA\cong\frg_1\oplus\frg_2$, where $\frg_1\cong\frg_2$, and the bilinear pairing is positive definite on $\frg_1$ and negative definite on $\frg_2$. The Chern-Simons Lagrangian then splits into two pieces of Chern-Simons type with a relative sign between the two Chern-Simons levels. Parity invariance can now be restored by postulating that under this transformation, the first Chern-Simons Lagrangian transforms into the second one and vice versa.

We also allow for superpotential terms, which we take to be of the following form:\footnote{We could also
have included terms involving the associated 3-product, but in the real case the ordinary 3-bracket already allows for marginal deformations. Additionally, one could introduce mass deformations of the form $\int\dd^{3|2}z\,R_{ij}(\Phi^i,\Phi^j)+\mbox{c.c.}$ but in 
this work we shall only be concerned with deformations that do not break conformal invariance already at the classical level.}
\begin{equation}\label{eq:S1Real}
\begin{aligned}
 S_1^R\ &=\ \int \dd^{3|2}z\,\left[R^{(1)}_{ijkl}(\Phi^l,[\Phi^i,\Phi^j,\Phi^k])
                +R^{(2)}_{ijkl}(\Phi^i,\Phi^j)(\Phi^k,\Phi^l)\right]\\
&\kern1cm+\int \dd^{3|2}\bar z\,\left[R^{ijkl}_{(1)}
     (\bar{\Phi}_l,[\bar{\Phi}_i,\bar{\Phi}_j,\bar{\Phi}_k])+R^{ijkl}_{(2)}(\Phib_i,\Phib_j)
      (\Phib_k,\Phib_l)\right],
\end{aligned} 
\end{equation}
where $\dd^{3|2}z:=\dd^3x\, \dd^2\theta$ and $\dd^{3|2}\bar z:=\dd^3x\, \dd^2\bth$ are the (anti)chiral
superspace measures. The symmetry properties of the 3-bracket and the pairing induces the following symmetry structures on the four-index parameters:
\begin{equation}
  R_{ijkl}^{(1)}\ =\ -R_{jikl}^{(1)}\ =\ -R_{ijlk}^{(1)}\ =\ R_{klij}^{(1)}\eand
  R_{ijkl}^{(2)}\ =\ R_{jikl}^{(2)}\ =\ R_{ijlk}^{(2)}\ =\ R_{klij}^{(2)}~.
\end{equation}
The couplings with upper indices are related to those with lower indices by complex conjugation,
\begin{equation}
 R_{ijkl}^{(1)}\ =\ (R^{ijkl}_{(1)})^*\eand
 R_{ijkl}^{(2)}\ =\ (R^{ijkl}_{(2)})^*~.
\end{equation}
The component form of the action $S_0^R+S_1^R$ is given in Appendix \ref{app:CFofA}

Note that the double trace term in the superpotential \eqref{eq:S1Real} corresponds to a double and a triple trace deformation in the potential. Note also that when discussing Feynman rules, the quartic terms $R^{(1)}$ and $R^{(2)}$
may be formally combined into one single vertex, cf.\ \eqref{eq:DefOfSymR} together with \
\eqref{eq:PPPP-Vertex} and \eqref{eq:PPPPbar-Vertex}.
Furthermore, the full supergauge transformations\footnote{after performing the integral over $t$} are given by
\begin{equation}
\begin{aligned}
\delta V\ &=\ \pounds_{-\frac{\di}{\sqrt{\kappa}}V}\big\{\Lambda-\bar\Lambda+\coth(\pounds_{-\frac{\di}{\sqrt{\kappa}}V})(\Lambda+\bar \Lambda)\big\}\ =\ \Lambda+\bar\Lambda-\tfrac{\di}{\sqrt{\kappa}}[V,\Lambda-\bar\Lambda]+\CO(1/\kappa)~,\\
\delta\Phi^i\ &=\ \tfrac{2\di}{\sqrt{\kappa}}\Lambda(\Phi^i)~,
\end{aligned}
\end{equation}
where $\pounds$ is the Lie-derivative $\pounds_{X}(Y)=\lsb X,Y\rsb$, $\coth(\pounds_{-\frac{i}{\sqrt{\kappa}}V})$ is defined via its series expansion and $\Lambda$ and $\bar\Lambda$ are the chiral and antichiral gauge parameters.

By construction, the above model has at least $\CN=2$ supersymmetry. Higher supersymmetry depends
on the underlying 3-algebra and the choices for the coefficients in the superpotential.
For instance, the 
original BLG model corresponds to 
\begin{equation}\label{eq:BLGmodel}
 \CA\ =\ A_4~,~~~N_f\ =\ 4~,~~~R_{ijkl}^{(1)}\ =\ \tfrac{\di}{4!\kappa}\eps_{ijkl}
 \eand R_{ijkl}^{(2)}\ =\ 0~,
\end{equation}
which yields the maximally supersymmetric theory with $\CN=8$ supersymmetry.

\subsection{Deformations of the superfield action in the Hermitian case}

In the Hermitian case which is based on Hermitian 3-algebras, the $\sSU(N_f)$ flavor multiplet will not be chiral, as discussed, e.g., in \cite{Minahan:2008hf}. Therefore, we have the set of chiral superfields $\Phi^i=(\Phi^1,\ldots,\Phi^{N_f})=(\Phi^m,\Phi^\mdt)$, but the $\sSU(N_f)$ flavor multiplet is formed by $(\Phi^m,\bar{\Phi}_\mdt)$. That is, we split the flavor index $i=1,\ldots,N_f$ into a pair $m,\mdt=1,\ldots,N_f/2$, where $\Phi^m$ and $\bar{\Phi}_{\mdt}$ will now be in the same representation of the gauge algebra whose carrier space is $\CA$.  Accordingly, we have to adjust the model here to read as 
\begin{equation}\label{eq:S0Complex}
\begin{aligned}
 S^H_0\ &=\ \di\sqrt{\kappa}\int 
  \dd^{3|4}z\int_0^1\dd t\,\lbr V,\bar{D}^\alpha\big(\de^{-\frac{2\di}{\sqrt{\kappa}} tV}
 D_\alpha \de^{\frac{2\di}{\sqrt{\kappa}} tV}\big)\rbr\\
 &\kern1cm+\int \dd^{3|4}z\,\big[(\Phi^m,\de^{-\frac{2\di}{\sqrt{\kappa}}V}\Phi^m)+
    (\Phib_\mdt,\de^{\frac{2\di}{\sqrt{\kappa}}V}\Phib_\mdt)\big]~.
\end{aligned}
\end{equation}
The unusual contraction of the flavor indices is due to the antilinearity of the third slot in the Hermitian 3-bracket and the first slot in the Hermitian pairing, respectively. The coupling constant $\kappa$ is again related to the Chern-Simons level $k$ via $\kappa=k/\pi$. 
We will allow for the following superpotential deformations, which preserve classical conformal invariance:
\begin{equation}\label{eq:S1Complex}
\begin{aligned}
 S_1^H\ &=\ \int \dd^{3|2}z\,\left[H_{mn\mdt\ndt}^{(1)}(\bar{\Phi}_\ndt,[\Phi^m,\Phi^n;\bar{\Phi}_\mdt]_\beta)+
      H_{mn\mdt\ndt}^{(2)}(\bar{\Phi}_\mdt,\Phi^m)(\bar{\Phi}_\ndt,\Phi^n)  \right]\\
 &\kern1cm+\int \dd^{3|2}\bar z\,\left[H^{\mdt\ndt mn}_{(1)}(\Phi^n,[\Phib_\mdt,\Phib_\ndt;\Phi^m]_\beta)+
      H^{\mdt\ndt mn}_{(2)}(\Phi^m,\bar{\Phi}_\mdt)(\Phi^n,\bar{\Phi}_\ndt)  \right]~,
\end{aligned} 
\end{equation}
where $[\cdot,\cdot;\cdot]_\beta$ was defined in \eqref{eq:BetaCommutator}.
The symmetry structure of the couplings here read as
\begin{equation}
H_{mn\mdt\ndt}^{(1)}\ =\ H_{nm\ndt\mdt}^{(1)}\eand
H_{mn\mdt\ndt}^{(2)}\ =\ H_{nm\ndt\mdt}^{(2)}~,
\end{equation}
and the relations of couplings with upper indices to the ones with lower indices are
\begin{equation}
 H_{mn\mdt\ndt}^{(1)}\ =\ -(H^{\ndt\mdt mn}_{(1)})^*\eand
 H_{mn\mdt\ndt}^{(2)}\ =\ (H^{\mdt\ndt mn}_{(2)})^*~.
\end{equation}
For the particular choice $\beta=0$, the $\beta$-3-bracket reduces to the Hermitian
3-bracket. In this case, the coupling $H^{(1)}_{mn\mdt\ndt}$ has the additional symmetry properties $H_{mn\mdt\ndt}^{(1)}=-H_{nm\mdt\ndt}^{(1)}=-H_{mn\ndt\mdt}^{(1)}$. Thus, for
$N_f=4$ it is of the form $H^{(1)}_{mn\mdt\ndt}\sim\varepsilon_{mn}\varepsilon_{\mdt\ndt}$.

Moreover, supergauge transformations in this case
are given by
\begin{equation}
\begin{aligned}
\delta V\ &=\ \pounds_{-\frac{\di}{\sqrt{\kappa}}V}\big\{\Lambda-\bar\Lambda+\coth(\pounds_{-\frac{\di}{\sqrt{\kappa}}V})(\Lambda+\bar \Lambda)\big\}\ =\ \Lambda+\bar\Lambda-\tfrac{\di}{\sqrt{\kappa}}[V,\Lambda-\bar\Lambda]+\CO(1/\kappa)~,\\
\delta\Phi^m\ &=\ \tfrac{2\di}{\sqrt{\kappa}}\Lambda(\Phi^m)\eand
\delta\Phib_\mdt\ =\ -\tfrac{2\di}{\sqrt{\kappa}}\bar\Lambda(\Phib_\mdt)~.
\end{aligned}
\end{equation}
Note again that the representation formed by $\Phi^\mdt$ is the complex conjugate representation of $\Phi^m$.
We refer to Appendix \ref{app:CFofA} for the component version of the above actions (for $\beta=0$).

The ABJM model as formulated in \cite{Bagger:2008se} is obtained by choosing $\CA=M_{\rm I_\alpha}^H(N)$ together with the couplings
\begin{equation}\label{eq:ABJMmodel}
  N_f\ =\ 4~,~~~\beta\ =\ 0~,~~~H_{mn\mdt\ndt}^{(1)}\ =\ \tfrac{1}{4\kappa}\varepsilon_{mn}\varepsilon_{\mdt\ndt}
  \eand H_{mn\mdt\ndt}^{(2)}\ =\ 0~,
\end{equation}
and putting $\alpha=1$ in \eqref{Herm_I}, one obtains exactly the ABJM model as written down, e.g., in \cite{Benna:2008zy}.

\section{Marginal deformations of the BLG and ABJM models}

All the superpotential terms introduced in the previous section are classically marginal. Recall that they were captured by parameters $R^{(\ell)}_{ijkl}$ and $H^{(\ell)}_{ijkl}$ for $\ell=1,2$. In the following, we will examine their behavior under quantization.

While the beta function of a pure three-dimensional WZ model is zero due to an argument 
analogous to \cite{Seiberg:1993vc}, the situation is different when we couple the model to a Chern-Simons action, cf.\ \cite{Gaiotto:2007qi}: In SYM theories it is possible to argue that the couplings in the superpotential do not renormalize by promoting the gauge coupling to a chiral superfield. The Chern-Simons level, however, is not a continuous parameter, and therefore this argument does not apply here.
Fortunately, it is known that the Chern-Simons level itself does not receive any quantum corrections, see e.g.\ \cite{Gates:1991qn,Kapustin:1994mt,DelCima:1997pb}, even if the model is coupled to arbitrary renormalizable matter theories.\footnote{cf.\ also the discussion in \cite{AlvarezGaume:1989wk,Chen:1992ee,Kao:1995gf} and more recently in \cite{Bedford:2008hn}.} It therefore suffices to study the beta function of the superpotential couplings. 

\subsection{Quantum action in the real case}\label{sec:QAandFR}

To discuss the renormalization of our models, we find it convenient to perform the
quantum computations directly in superspace. For textbook treatments of the supergraph formalism
in the SYM case in four dimensions, which is very similar to our discussion below, we refer e.g.\ to \cite{Gates:1983nr,Buchbinder:1998qv}.

Let us start from the action \eqref{eq:S0Real} in the real setting. 
The Hermitian case of \eqref{eq:S0Complex} is treated 
analogously, and we will discuss the differences in Section \ref{eq:BetaHC} We shall suppress the superscript $R$ in the following.

First, let us expand \eqref{eq:S0Real}
in powers of $V$. For our purposes, it will be enough to keep terms
only up to $\CO(V^3)$,
\begin{equation}
 \begin{aligned}
  S_0\ &=\ \int \dd^{3|4}z\left[\di\lbr V,\bar D_\alpha D^\alpha V\rbr+\tfrac{2}{3\sqrt{\kappa}}
  \lbr V,\lsb D_\alpha V,\Db^\alpha V\rsb\rbr+(\Phib_i,\Phi^i)+\tfrac{-2\di}{\sqrt{\kappa}}(\Phib_i,V\Phi^i)\right.\\
    &\kern1cm\left.
         +\tfrac{1}{2!}\big(\tfrac{-2\di}{\sqrt{\kappa}}\big)^2(\Phib_i,V^2\Phi^i)
         +\tfrac{1}{3!}\big(\tfrac{-2\di}{\sqrt{\kappa}}\big)^3(\Phib_i,V^3\Phi^i)+\CO(V^4)\right].
 \end{aligned}
\end{equation}
Here and in the following, the bracket $\lsb\cdot,\cdot\rsb$ denotes the supercommutator, i.e.\ an anticommutator if the Gra\ss mann parity of both arguments is odd and a commutator otherwise.

To quantize this action, we adopt a supersymmetric Landau gauge as done e.g.\ in \cite{Avdeev:1991za,Gates:1991qn}. The corresponding gauge fixing term reads as\footnote{Alternatively, we could have introduced the usual
gauge fixing Lagrangian $\CL_{\rm gf}\sim \frac1\xi\lbr V,D^2\Db^2V+\Db^2 D^2V\rbr$ at the cost of having
a dimensionful gauge parameter $\xi$; in fact, since $V$ is dimensionless, $\xi$ is of mass-dimension 1.
As a consequence, the corresponding gluon propagator has a bad IR behavior for
$\xi\neq0$. However, for $\xi=0$ the propagator is the same as the one
given in \eqref{eq:GluonProp} for $\alpha\beta\to0$.}
\begin{equation}\label{eq:gaugefixingaction}
 S_{\rm gf}\ =\ \int \dd^{3|4}z\,\lbr V, \{\alpha^{-1}(D^2+\Db^2)-\di \beta^{-1}(D^2-\Db^2)\}V \rbr~, 
\end{equation}
where we take the limit $\alpha\beta\to0$. Here, $\alpha$ and $\beta$ are dimensionless parameters and
$D^2:=D^\alpha D_\alpha$ and $\Db^2:=\Db_\alpha \Db^\alpha$. Accordingly, the
Faddeev-Popov action is
\begin{equation}\label{eq:ghostaction}
 S_{\rm gh}\ =\ \int \dd^{3|4}z\,\lbr b-\bar b, \pounds_{-\frac{i}{\sqrt{\kappa}}V}\big\{c-\bar c+\coth\big(\pounds_{-\frac{i}{\sqrt{\kappa}}V}\big)(c+\bar c)\big\} \rbr ~,
\end{equation}
where the $c$ are the ghosts while the $b$ are the antighosts; these are (anti)chiral
superfields.

As one may check, $S_{\rm gf}+S_{\rm gh}$ is invariant under the following BRST transformation laws:
\begin{equation}
 \begin{aligned}
  \delta_{BRST} V \ &=\ \tfrac{\di\sqrt{\kappa}}{2}\,\eta\, \pounds_{-\frac{i}{\sqrt{\kappa}}V}\big\{c-\bar c+\coth\big(\pounds_{-\frac{i}{\sqrt{\kappa}}V}\big)(c+\bar c)\big\}~,\\
\delta_{BRST} c\ &=\ -\eta\,c^2\eand
  \delta_{BRST} \bar c\ =\ \eta\,\bar c^2~,\\
  \delta_{BRST} b\ &=\ -\di\sqrt{\kappa}\,\eta\,(\alpha^{-1}-\di \beta^{-1})\Db^2V~,\\
  \delta_{BRST} \bar b\ &=\ \di\sqrt{\kappa}\,\eta\,(\alpha^{-1}+\di\beta^{-1})D^2V~,
 \end{aligned}
\end{equation}
where $\eta$ is some anticommuting parameter.

For our purposes, we will need $S_{\rm gh}$ only to $\CO(V^1)$,
\begin{equation}
 S_{\rm gh}\ =\ \int \dd^{3|4}z\,\left[-\lbr\bar b,c\rbr-\lbr\bar c,b\rbr-\tfrac{\di}{\sqrt{\kappa}}
   \lbr b-\bar b,\lsb V,c-\bar c\rsb\rbr+\CO(V^2)\right].
\end{equation}
The full quantum action is then given by
\begin{equation}\label{eq:QuantumAction}
 S_q\ =\ S_0+S_1+S_{\rm gf}+S_{\rm gh}~.
\end{equation}

In order to have a compact form of the Feynman rules, we use capital Roman letters $A,B,\ldots=1,\ldots,\operatorname{dim}\frg_\CA$ to denote gauge algebra indices. For this, it is important to note that there is a priori no bijection between pairs of indices $ab$ denoting elements of $\Lambda^2\CA$ and an index $A$ corresponding to an element of $\frg_\CA$. This is due to the fact that $\delta:\Lambda^2 \CA\rightarrow \frg_\CA$ is not injective in general (with an exception being the case of the real 3-algebra $A_4$). This point has to be carefully taken into account in all the calculations in the following. 

In terms of the gauge algebra indices, the invariant form $\lbr\cdot,\cdot\rbr$ on $\frg_{\CA}$ is simply given by
\begin{equation}
 \lbr X,Y\rbr\ =\ X^{ab} Y^{bc} f_{abcd}\ =:\ X^A Y^B G_{AB}~,\ewith G_{AB}\ =\ G_{BA}~.
\end{equation}
We assume that $G_{AB}$ has an inverse denoted by $G^{AB}$ with $G_{AC}G^{CB}={\delta_A}^B$. Note that the identification $G_{AB}=f_{abcd}$ holds only if $\delta$ is a bijection (as is the case for $\CA=A_4$). The structure constants of $\frg_{\CA}$ are denoted by ${F_{AB}}^C$. In interactions like the 3-gluon vertex, the quantity $F_{ABC}:= {F_{AB}}^D G_{DC}$ will appear. Due to $ad$-invariance
of $\lbr\cdot,\cdot\rbr$, $F_{ABC}$ is totally antisymmetric in $ABC$. Moreover, we will use multi-indices $I=ia$ combining flavor and 3-algebra indices whenever convenient. For example, vertices like
\begin{subequations}
\begin{equation}
 (\Phib_i,V(\Phi^i))\ =\ V^{ab}\Phi^{ic}\Phib_i^{d} f_{abcd}\ =\ V^{ab}\Phi^{jc}\Phib_{id}
 {f_{abc}}^d\delta_i^{~j}~,
 \ewith\Phib_{ia}\ :=\ h_{ab}\Phib_i^b
\end{equation}
that appear in the expansion of $(\bar{\Phi}_i,\de^{-\frac{2\di}{\sqrt{\kappa}}V}\Phi^i)$, will be written
as
\begin{equation}
 (\Phib_i,V(\Phi^i))\ =\ \Phi^I V^A {T_{AI}}^J \Phib_J~,
\end{equation}
where
\begin{equation}
 [T_A,T_B]\ =\ {F_{AB}}^C T_C~.
\end{equation}
\end{subequations}
We stress again that the identification ${T_{AI}}^J={f_{abc}}^d\delta_i^{~j}$ works only if $\delta:\Lambda^2\CA\rightarrow \frg_\CA$ is a bijection.

\subsection{Feynman rules}

We have now all the necessary ingredients to write down the momentum space Feynman rules for our theory ($\partial_\mu\mapsto-\di p_\mu$). 

\bigskip
\noindent{\it \underline{Propagators:}}
\bigskip

\noindent 
The propagators are found to be\footnote{Here and in the sequel, we make no notational distinction
between a position space field and its momentum space version (after Fourier transform).}
\begin{subequations}\label{eq:propagators}
 \begin{eqnarray}
\kern-.8cm
\begin{picture}(80,20)(-5,0)
   \SetScale{1}
   \put(0,0){
    \Photon(0,2)(65,2){3}{12}
    \Text(65,12)[l]{$B$}
    \Text(65,-10)[l]{$\theta_2$}
    \Text(2,12)[r]{$A$}
    \Text(2,-10)[r]{$\theta_1$}
    \Text(34,10)[]{$\longrightarrow$}
    \Text(36,15)[]{$p$}
   }
  \end{picture}
 \kern8pt:\kern8pt
\langle V^A(-p,\theta_1) V^B(p,\theta_2)\rangle\! &=&\notag\\[12pt]
  &&\kern-8.5cm=\ -\frac{\di}{4p^2}G^{AB}\left[\Db_\alpha D^\alpha-\tfrac{\di}{4}\tfrac{\alpha^2\beta^2}{\alpha^2+\beta^2}
        \left\{\alpha^{-1}(D^2+\Db^2)-
 \di \beta^{-1}(D^2-\Db^2)\right\}\right]\delta^{(4)}(\theta_1-\theta_2)~,\label{eq:GluonProp}\\[12pt]
\kern-.8cm
\begin{picture}(80,20)(-4,0)
   \SetScale{1}
   \put(0,0){
    \ArrowLine(0,2)(65,2)    
    \Text(65,12)[l]{$J$}
    \Text(65,-10)[l]{$\theta_2$}
    \Text(2,12)[r]{$I$}
    \Text(2,-10)[r]{$\theta_1$}
    \Text(34,10)[]{$\longrightarrow$}
    \Text(36,15)[]{$p$}
   }
  \end{picture}
 \kern8pt:\kern8pt
 \langle \Phi^{I}(-p,\theta_1)\Phib_J(p,\theta_2)\rangle\! &=&\! -\frac{\di}{p^2}{\delta^I}_J   \delta^{(4)}(\theta_1-\theta_2)~,\\[12pt]
\kern-.8cm
\begin{picture}(80,20)(0,0)
   \SetScale{1}
   \put(0,0){
    \DashArrowLine(0,2)(65,2){4}    
    \Text(65,12)[l]{$B$}
    \Text(65,-10)[l]{$\theta_2$}
    \Text(2,12)[r]{$A$}
    \Text(2,-10)[r]{$\theta_1$}
    \Text(34,10)[]{$\longrightarrow$}
    \Text(36,15)[]{$p$}
   }
  \end{picture}
 \kern8pt:\kern8pt \langle c^A(-p,\theta_1)\bar b^B(p,\theta_2)\rangle\! &=&\! \frac{\di}{p^2}G^{AB}\delta^{(4)}(\theta_1-\theta_2)~,\\[12pt]
\kern-.8cm
\begin{picture}(80,20)(0,0)
   \SetScale{1}
   \put(0,0){
    \DashArrowLine(0,2)(65,2){4}    
    \Text(65,12)[l]{$B$}
    \Text(65,-10)[l]{$\theta_2$}
    \Text(2,12)[r]{$A$}
    \Text(2,-10)[r]{$\theta_1$}
    \Text(34,10)[]{$\longrightarrow$}
    \Text(36,15)[]{$p$}
   }
  \end{picture}
 \kern8pt:\kern8pt
 \langle b^A(-p,\theta_1)\bar c^B(p,\theta_2)\rangle\! &=&\! 
   \frac{\di}{p^2}G^{AB}\delta^{(4)}(\theta_1-\theta_2)~,
 \end{eqnarray}
\end{subequations}
where all derivatives are understood to depend on $p$ and to act on $\theta_1$.\footnote{Note that we make no pictorial distinction between
$\langle c\bar b\rangle $ and $\langle b\bar c\rangle $.}
Here, we suppressed the usual $\di\varepsilon$-prescription of the poles. 
As already indicated, in this work we will use Landau gauge with $\alpha\beta\to0$. We shall also use the convention $\Db_\alpha D^\alpha = \Db D$.

\bigskip
\noindent{\it \underline{Vertices:}}
\bigskip

\noindent
Vertices can be read off directly from the action \eqref{eq:QuantumAction}, and for the reader's
convenience we have summarized them in Appendix \ref{app:vertices} As for SYM theory in superspace language, there is one additional feature that for each chiral or
antichiral line leaving a vertex there is a factor of $-\frac14 \Db^2$ or $-\frac14 D^2$
acting on the corresponding propagator. However, for purely chiral or antichiral
vertices that come from the superpotential, we omit one factor of $-\frac14 \Db^2$ or $-\frac14 D^2$
corresponding to one internal line, i.e.\ a vertex with $n$ internal lines attached
carries $n-1$ derivative factors.

\bigskip
\noindent{\it \underline{Integration, symmetry factors and regularization:}}
\bigskip

\noindent
First, there are the usual loop-momentum integrals $\int\frac{\dd^3p}{(2\pi)^3}$ for each loop and momentum conserving delta functions. Second, we integrate over $\dd^4\theta$ at each vertex. Finally, the usual symmetry factors associated with the diagram have to be taken into account.

Our regularization prescription is as follows: We will perform all manipulations of the formul\ae{} in $D=3$, $\CN=2$
superspace and only compute the final loop-momentum integrals in dimensional regularization. This prescription corresponds to dimensional reduction \cite{Siegel:1979wq}, a procedure, which is known to be valid at least up to two loop order \cite{Chen:1992ee}.

\subsection{Powercounting}

Before performing the calculation, it is useful to look at the superficial
degree of divergence $\delta(\Gamma)$ of some diagram $\Gamma$.

With the given Feynman rules, the gluon propagator $\langle VV\rangle$ scales as $1/p$ for large momenta
while the propagators for the matter $\langle\Phi\Phib\rangle$ and the ghosts 
$\langle c\bar b\rangle$ and $\langle b\bar c\rangle$ go like $1/p^2$.  The $V^{n}$ vertex scales as $D\Db\sim p$, each vertex of type $\Phi V^n\Phib$ goes like $D^2\Db^2\sim p^2$ and the $\Phi^4$ and $\Phib^4$ vertices behave as $\Db^6\sim p^3$ and $D^6\sim p^3$, respectively. Any ghost/gluon
interaction goes like $D^2\Db^2\sim p^2$. Then each external chiral or antichiral line (matter and ghost lines) goes
like $1/\bD^2\sim 1/p$ or $1/D^2\sim 1/p$. Finally, as in SYM theory in four dimensions \cite{Grisaru:1979wc}, for each loop one may reduce all the $\dd^4\theta$-integrals to just a single one by partially integrating the $D$- and $\Db$-derivatives, hence leaving each loop-momentum integral to behave as $\dd^3 p/D^2\Db^2\sim p$.

Altogether, the superficial degree of divergence is thus given by
\begin{equation}
 \delta(\Gamma)\ =\ V_g+2V_{cg}+3V_c-I_g-2I_c+L-E_c~,
\end{equation}
where $V_g$ is the number of purely gluonic vertices, $V_{cg}$ the number of matter/gluon and ghost/gluon interactions and $V_c$ is the number of purely chiral vertices of $\Gamma$. Then, $I_g$ is the number
of internal gluon lines, $I_c$ is the number of ghost and matter lines and $E_c$ is the number of external ghost and matter lines. Finally,
$L$ is the number of loops. 

Using the formul\ae{}
\begin{equation}
 L\ =\ I-V+1\ =\ I_g+I_c-V_g-V_{cg}-V_c+1\eand E_c+2I_c \ =\ 2V_{cg}+4V_c~,
\end{equation}
we eventually arrive at
\begin{equation}\label{eq:SDofD}
 \delta(\Gamma) \ =\ \tfrac{1}{2}(2-E_c)~.
\end{equation}
Comparing this with the result of SYM theory in four dimensions, \cite{Grisaru:1979wc}, we conclude that
$\delta_{SCS}=\frac12\delta_{SYM}$.

Equation \eqref{eq:SDofD} then tells us that all diagrams with more than two external chiral lines are superficially
convergent. Notice that \eqref{eq:SDofD} can be refined further. When partially integrating the supercovariant derivatives some of them will get transferred to external lines (when, e.g., computing the wave function renormalization of the vector superfield or the renormalization of the superpotential). 
If we let $N_D$ be the number of $D$- and $\Db$-derivatives that are transferred to external lines, then
the superficial degree of divergence is given by
\begin{equation}\label{eq:SDofD2}
 \delta(\Gamma) \ =\ \tfrac{1}{2}(2-E_c-N_D)~.
\end{equation}

\subsection{Two-loop renormalization in the real case}

Let us now come to the computation of the beta functions $\beta_{ijkl}^{(\ell)}$
for the couplings $R_{ijkl}^{(\ell)}$  with $\ell=1,2$. Upon
rescaling $\Phi^i_0=(Z^{1/2})_j^{~i}\Phi^j$, where the subscript `0' refers to the bare quantities, we find
\begin{equation}
 R_{0\,ijkl}^{(\ell)}\ =\ (Z^{-1/2})_i^{~i'}\cdots(Z^{-1/2})_l^{~l'}{Z^{(\ell)}_{i'j'k'l'}}\,\!^{i''j''k''l''} 
 R_{i''j''k''l''}^{(\ell)}~,
\end{equation}
and hence
\begin{subequations}
\begin{equation}
\begin{aligned}
 \beta_{ijkl}^{(\ell)}\ &=\ 
      {(Z^{-1})_{ijkl}}^{i'j'k'l'}\gamma_{i'}^{~m}{Z^{(\ell)}_{mj'k'l'}}\,\!^{i''j''k''l''}R_{i''j''k''l''}^{(\ell)}\\
         &\kern1cm +\cdots+         
       {(Z^{-1})_{ijkl}}^{i'j'k'l'}\gamma_{l'}^{~m}{Z^{(\ell)}_{i'j'k'm}}\,\!^{i''j''k''l''}
       R_{i''j''k''l''}^{(\ell)}\\
       &\kern2cm+{((Z^{(\ell)})^{-1})_{ijkl}}\,\!^{i'j'k'l'}
 \frac{\dd {Z^{(\ell)}_{i'j'k'l'}}\,\!^{i''j''k''l''}}{\dd\log \mu}
R_{i''j''k''l''}^{(\ell)}~,
\end{aligned}
\end{equation}
where
\begin{equation}
 \gamma_i^{~j}\ =\ (Z^{-1/2})_i^{~k}\frac{\dd (Z^{1/2})_k^{~j}}{\dd\log \mu}\ =\ 
                   \frac12 \frac{\dd(\log Z)_i^{~j}}{\dd\log \mu}
\end{equation}
\end{subequations}
denotes the anomalous dimension of the field $\Phi^i$ and ${Z^{(\ell)}_{ijkl}}\,\!^{i'j'k'l'}$ 
is the renormalization of the quartic vertex $\ell$.

\begin{figure}[h]
\begin{center}
\begin{picture}(100,110)(60,-10)
   \SetScale{1}
   \put(0,0){
    \Vertex(28,50){1}
    \Vertex(72,50){1}
    \PhotonArc(50,50)(22,0,383){3}{25.5}
    \ArrowLine(0,50)(28,50)
    \ArrowLine(28,50)(72,50)
    \ArrowLine(72,50)(100,50)
    \LongArrowArc(50,58)(22,60,120)
    \LongArrowArc(50,42)(22,240,300)
    \Text(105,50)[l]{{\footnotesize $\Phib_J(p,\theta_2)$}}
    \Text(-47,50)[l]{{\footnotesize $\Phi^I(-p,\theta_1)$}}
    \Text(12,42)[l]{{\footnotesize $p$}}
    \Text(47,42)[l]{{\footnotesize $k$}}
    \Text(84,42)[l]{{\footnotesize $p$}}
    \Text(32,87)[l]{{\footnotesize $k+l-p$}}
    \Text(47,13)[l]{{\footnotesize $l$}}
    \Text(44,0)[l]{{\footnotesize $(a)$}}
   }
\end{picture}
\begin{picture}(100,110)(-60,-10)
   \SetScale{1}
   \put(0,0){
    \Vertex(28,50){1}
    \Vertex(72,50){1}
    \ArrowArc(50,50)(22,0,180)
    \ArrowArcn(50,50)(22,360,180)
    \ArrowLine(0,50)(28,50)
    \ArrowLine(72,50)(28,50)
    \ArrowLine(72,50)(100,50)
    \Text(105,50)[l]{{\footnotesize $\Phib_J(p,\theta_2)$}}
    \Text(-47,50)[l]{{\footnotesize $\Phi^I(-p,\theta_1)$}}
    \Text(12,42)[l]{{\footnotesize $p$}}
    \Text(47,42)[l]{{\footnotesize $k$}}
    \Text(84,42)[l]{{\footnotesize $p$}}
    \Text(28,80)[l]{{\footnotesize $-k-l-p$}}
    \Text(47,18)[l]{{\footnotesize $l$}}
    \Text(47,0)[l]{{\footnotesize $(b)$}}
   }
\end{picture}

\begin{picture}(100,110)(0,-10)
   \SetScale{1}
   \put(0,0){
    \Vertex(50,30){1}
    \PhotonArc(50,55)(22,120,420){3}{18.5}
    \ArrowLine(0,30)(50,30)
    \ArrowLine(50,30)(100,30)
    \GOval(50,74)(11,11)(0){0.7}
    \LongArrowArc(50,70)(22,60,120)
    \Text(105,30)[l]{{\footnotesize $\Phib_J(p,\theta_1)$}}
    \Text(-45,30)[l]{{\footnotesize $\Phi^I(-p,\theta_1)$}}
    \Text(22,22)[l]{{\footnotesize $p$}}
    \Text(72,22)[l]{{\footnotesize $p$}}
    \Text(48,100)[l]{{\footnotesize $k$}}
    \Text(44,0)[l]{{\footnotesize $(c)$}}
   }
\end{picture}
\end{center}
\vspace*{-15pt}
\caption{Logarithmically divergent diagrams that contribute to $Z_i^{~j}$.}\label{fig:ZDiagrams}
\end{figure}
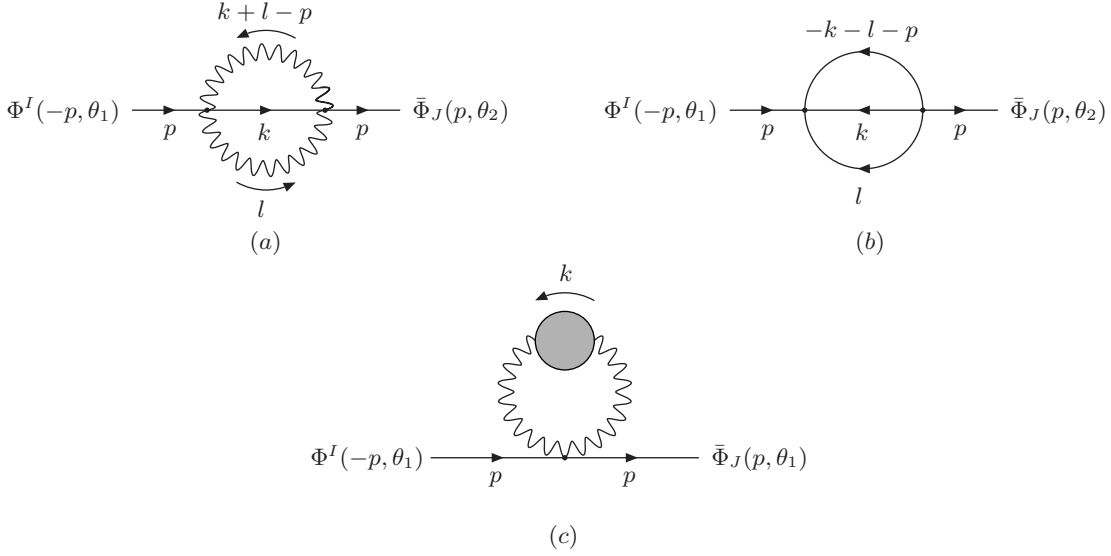

To compute $\beta_{ijkl}^{(\ell)}$, we emphasize that there is no one-loop renormalization, as there are no Feynman diagrams which could potentially contribute. Note that Lemma 3 of \cite{Gates:1991qn} is very helpful here, as it immediately rules out contributions from large classes of diagrams. The first non-trivial result is found at two loops. From the discussion in the previous section, we conclude that all four-point functions are superficially convergent and indeed, by inspecting
all two-loop four-point diagrams of types $\langle\Phi\Phi\Phi\Phi\rangle$ and
$\langle\Phib\Phib\Phib\Phib\rangle$ explicitly, one realizes that they all are convergent: There is a single such diagram potentially contributing (the two-loop gluon correction to the vertex), which is, however, convergent. We are therefore left with
\begin{equation}\label{eq:BetaFunction}
 \beta_{ijkl}^{(\ell)}\ =\ \gamma_i^{~m}R_{mjkl}^{(\ell)}+\cdots+\gamma_l^{~m}R_{ijkm}^{(\ell)}~.
\end{equation}
Moreover, there are only three diagrams that contribute to $\gamma_i^{~j}$ and they are displayed
in Fig.\ \ref{fig:ZDiagrams}. All other diagrams either vanish by supersymmetry or by their respective color structure or they are simply finite. 

Furthermore, it will be helpful to introduce the following operators:
\begin{equation}\label{eq:ops123}
\begin{aligned}
 (\CCO_1)_I^{~J}\ &:=\ G^{AB}{T_{AI}}^K{T_{BK}}^L G^{CD}{T_{CL}}^M{T_{DM}}^J~,\\
 (\CCO_2)_I^{~J}\ &:=\ \tfrac14 G^{AC}G^{BD} {F_{AB}}^E {F_{CD}}^F {T_{EI}}^{~K} {T_{FK}}^{~J}~,\\
 (\CCO_3)_I^{~J}\ &:=\ \tfrac{1}{N_f}{T_{AK}}^L {T_{BL}}^K G^{AC} G^{BD}{T_{CI}}^M {T_{DM}}^J~,
\end{aligned}
\end{equation}
and one can show that they commute with all $T_A$. However, the $T_A$ need not form an irreducible representation of $\frg_\CA$ in general, so Schur's lemma cannot be applied directly. Nevertheless, it turns out that for the 3-algebras we are interested in, i.e.\ $A_4$ and the class $M_{\rm III_{\alpha,\beta}}^R(N)$ and also
later for $M_{\rm I_{\alpha}}^H(N)$, 
the operators \eqref{eq:ops123} are indeed proportional to the identity. In these cases, we define
\begin{equation}
 (\CCO_1)_I^{~J}\ =:\ k_1^2\delta_I^{~J}~,~~~
 (\CCO_2)_I^{~J}\ =:\ k_2\delta_I^{~J}~,~~~
 (\CCO_3)_I^{~J}\ =:\ k_3 \delta_I^{~J}~.
\end{equation}
The explicit values of $k_1$, $k_2$ and $k_3$ for the various matrix representations are listed in Appendix \ref{app:UsefulFormulae} To be concise, we will give all our formul\ae{} using these constants in 
the following.

Let us start from diagram \ref{fig:ZDiagrams}a).
Using the Feynman rules listed in the previous section
and in Appendix \ref{app:vertices}, this diagram is given by the following integral:
\begin{equation}\label{eq:DiagramAa}
 \begin{aligned}
  \Sigma^{(a)}\ &=\ -\frac{\di}{16\cdot2\kappa^2}\big[k_2+k_1^2\big]\int\frac{\dd^3p}{(2\pi)^3}\frac{\dd^3k}{(2\pi)^3}\frac{\dd^3l}{(2\pi)^3}\,\dd^4\theta_1\dd^4\theta_2\,
  \Phi^I(-p,\theta_1)\Phib_I(p,\theta_2)\\
  &\kern4cm\times\frac{[D^2\Db^2(k,\theta_1)\delta_{21}][\Db D(k+l-p,\theta_2)\delta_{12}][\Db D(l,\theta_1)\delta_{12}]}{k^2 l^2 (k+l-p)^2}~,
 \end{aligned}
\end{equation}
where $\delta_{12}:=\delta^{(4)}(\theta_1-\theta_2)$; the $1/2$ is the symmetry factor. We arrive at this expression after making use of the {\it transfer rule}
\begin{equation}
 D(p,\theta_1)\delta_{12}\ =\ -D(-p,\theta_2)\delta_{12}~,
\end{equation}
where $D$ represents both, $D$ and $\Db$. 

Integrating by parts and by employing the $D$-algebra $\{D,\Db\}\sim p$, $\{D,D\}=0$ and $\{\Db,\Db\}=0$,
the integral \eqref{eq:DiagramAa} simplifies to
\begin{eqnarray}
 \Sigma^{(a)}\! &=&\! -\frac{4\di}{\kappa^2}\big[k_2+k_1^2\big]\int\frac{\dd^3p}{(2\pi)^3}\,\dd^4\theta\, 
              \Phi^I(-p,\theta)\Phib_I(p,\theta)\underbrace{\int\frac{\dd^3k}{(2\pi)^3}\frac{\dd^3l}{(2\pi)^3}\,
                \frac{1}{k^2l^2(k+l-p)^2}}_{=\ -\frac{\log\Lambda}{16\pi^2}}\notag\\
          &=&\! \frac{\di}{4\pi^2\kappa^2}\big[k_2+k_1^2\big]
             \log\Lambda\,\int\frac{\dd^3p}{(2\pi)^3}\, \dd^4\theta\,
              (\Phib_i(p,\theta),\Phi^i(-p,\theta))~.
\end{eqnarray}
Thus, the contribution of $\Sigma^{(a)}$ to $Z_i^{~j}$ is
\begin{equation}\label{eq:Za}
 \mbox{(a) :}~~~~\delta Z_i^{~j}\ =\ -\frac{\log\Lambda}{4\pi^2\kappa^2}\big[k_2+k_1^2\big]\delta_i^{~j}~.
\end{equation}

In a very similar manner, we find the contribution coming from diagram \ref{fig:ZDiagrams}b) to be
\begin{equation}\label{eq:Zb}
\begin{aligned}
 \mbox{(b) :}~~~~\delta Z_i^{~j}\ &=\ -\frac{2\log\Lambda}{\pi^2}
  \left[R_{iklm}^{(1)}\big(-c_3R^{jklm}_{(1)}+2c_2 R^{jmlk}_{(1)}+2c_1R^{jmlk}_{(2)}\big)\right.\\
&\kern2.5cm\left.+\ R_{iklm}^{(2)}\big(d\,R^{jklm}_{(2)}+2 R^{jmlk}_{(2)}+2c_1R^{jmlk}_{(1)}\big)
       \right]~,
\end{aligned}
\end{equation}
where $c_1$, $c_2$ and $c_3$ are the three ``Casimirs'' of $\CA$ that are given by 
\begin{equation}
 {f_{ac}}^{cb}\ =\ c_1{\delta_a}^b~,~~~
 f_{acde}f^{bedc}\ =\ c_2{\delta_a}^b\eand
 f_{acde}f^{bcde}\ =\ -c_3\delta_a^{~b}~
\end{equation}
and $d=\operatorname{dim}\CA$ is the dimension of the 3-algebra. These relations follow from the fundamental identity. We refer to Appendix 
\ref{app:UsefulFormulae}, where we list $c_1$, $c_2$ and $c_3$
for the matrix representation $M_{\rm III_{\alpha,\beta}}^R(N)$.

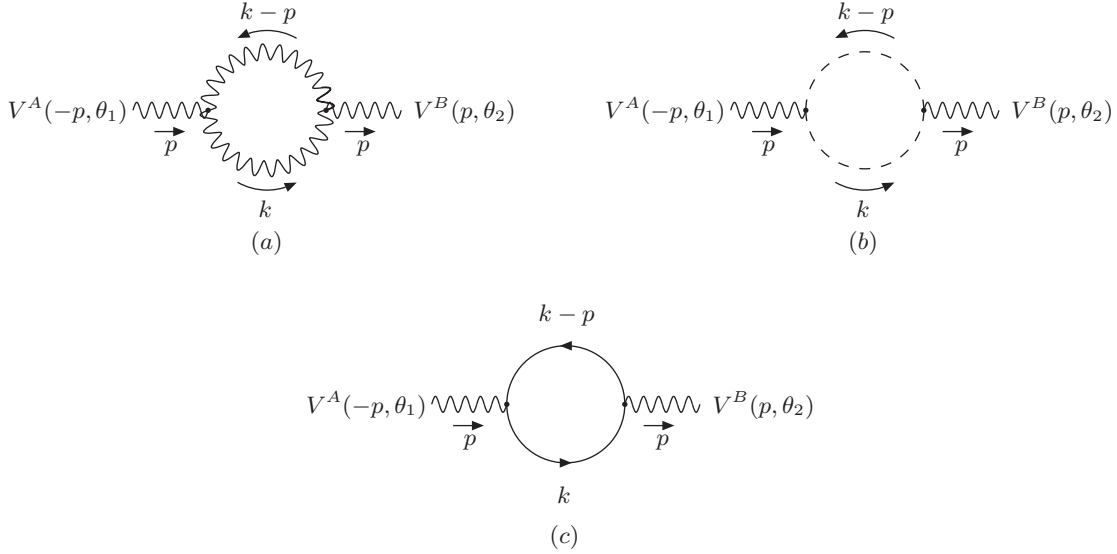
\begin{figure}[h]
\begin{center}
\begin{picture}(100,110)(60,-10)
   \SetScale{1}
   \put(0,0){
    \Vertex(28,50){1}
    \Vertex(72,50){1}
    \PhotonArc(50,50)(22,0,383){3}{25.5}
    \Photon(0,50)(28,50){3}{5}
    \Photon(72,50)(100,50){3}{5}
    \LongArrowArc(50,58)(22,60,120)
    \LongArrowArc(50,42)(22,240,300)
    \LongArrow(8,42)(18,42)
    \LongArrow(79,42)(89,42)
    \Text(105,50)[l]{{\footnotesize $V^B(p,\theta_2)$}}
    \Text(-47,50)[l]{{\footnotesize $V^A(-p,\theta_1)$}}
    \Text(12,36)[l]{{\footnotesize $p$}}
    \Text(84,36)[l]{{\footnotesize $p$}}
    \Text(40,87)[l]{{\footnotesize $k-p$}}
    \Text(47,13)[l]{{\footnotesize $k$}}
    \Text(44,0)[l]{{\footnotesize $(a)$}}
   }
\end{picture}
\begin{picture}(100,110)(-60,-10)
   \SetScale{1}
   \put(0,0){
    \Vertex(28,50){1}
    \Vertex(72,50){1}
    \DashCArc(50,50)(22,0,180){4}
    \DashCArc(50,50)(22,180,360){4}
    \Photon(0,50)(28,50){3}{5}
    \Photon(72,50)(100,50){3}{5}
    \LongArrowArc(50,58)(22,60,120)
    \LongArrowArc(50,42)(22,240,300)
    \LongArrow(8,42)(18,42)
    \LongArrow(79,42)(89,42)
    \Text(105,50)[l]{{\footnotesize $V^B(p,\theta_2)$}}
    \Text(-47,50)[l]{{\footnotesize $V^A(-p,\theta_1)$}}
    \Text(12,36)[l]{{\footnotesize $p$}}
    \Text(84,36)[l]{{\footnotesize $p$}}
    \Text(40,87)[l]{{\footnotesize $k-p$}}
    \Text(47,13)[l]{{\footnotesize $k$}}
    \Text(44,0)[l]{{\footnotesize $(b)$}}
   }
\end{picture}

\begin{picture}(100,110)(0,-10)
 \SetScale{1}
   \put(0,0){
    \Vertex(28,50){1}
    \Vertex(72,50){1}
    \ArrowArc(50,50)(22,0,180)
    \ArrowArc(50,50)(22,180,360)
    \Photon(0,50)(28,50){3}{5}
    \Photon(72,50)(100,50){3}{5}
    \LongArrow(8,42)(18,42)
    \LongArrow(79,42)(89,42)
    \Text(105,50)[l]{{\footnotesize $V^B(p,\theta_2)$}}
    \Text(-47,50)[l]{{\footnotesize $V^A(-p,\theta_1)$}}
    \Text(12,36)[l]{{\footnotesize $p$}}
    \Text(84,36)[l]{{\footnotesize $p$}}
    \Text(40,84)[l]{{\footnotesize $k-p$}}
    \Text(47,16)[l]{{\footnotesize $k$}}
    \Text(44,0)[l]{{\footnotesize $(c)$}}
   }
\end{picture}
\end{center}
\vspace*{-15pt}
\caption{One-loop diagrams that contribute to the gluon self-energy $\Pi$; they are all finite. The ghost diagram (b) represents all four ghost contributions.}\label{fig:GluonSelfEnergy}
\end{figure}

Finally, we need to find the contribution coming from diagram \ref{fig:ZDiagrams}c). To compute this
diagram, it is useful to perform the calculation in two steps. Let us first compute the one-loop
contributions to the gluon self-energy $\Pi$. For this, we introduce the usual superspin projectors $\CCP_0$ and $\CCP_{1/2}$,
\begin{equation}
 \CCP_0\ :=\ -\frac{1}{16p^2}\left[D^2\Db^2+\Db^2 D^2\right]\eand \CCP_{1/2}\ :=\ \frac{1}{8p^2} D^\alpha \Db^2 D_\alpha~,
\end{equation}
which obey
\begin{equation}
 \CCP_0^2\ =\ \CCP_0~,~~~\CCP_{1/2}^2\ =\ \CCP_{1/2}\eand\CCP_0+\CCP_{1/2}\ =\ 1~.
\end{equation}
With these, the relevant diagrams displayed in Fig.\ \ref{fig:GluonSelfEnergy} contribute according to
\begin{subequations}\label{eq:GSE}
 \begin{eqnarray}
   \Pi^{(a)}\! &=&\! -\frac{\di}{8\kappa}{F_{AC}}^D{F_{BD}}^C\int\frac{\dd^3 p}{(2\pi)^3}\,
    \dd^4\theta\,V^A(-p,\theta)\, p\, \CCP_0 V^B(p,\theta)~,\\
   \Pi^{(b)}\! &=&\! \frac{\di}{8\kappa}{F_{AC}}^D{F_{BD}}^C\int\frac{\dd^3 p}{(2\pi)^3}\,\dd^4\theta\, V^A(-p,\theta)\, p\, (\CCP_{1/2}+\CCP_0) V^B(p,\theta)~,\\
   \Pi^{(c)}\! &=&\! -\frac{\di}{4\kappa}{T_{AI}}^J{T_{BJ}}^I\int\frac{\dd^3 p}{(2\pi)^3}\,\dd^4\theta\, V^A(-p,\theta)\, p\, \CCP_{1/2}V^B(p,\theta)~,
 \end{eqnarray}
\end{subequations}
as follows by using the Feynman rules listed in Section \ref{sec:QAandFR} and in Appendix \ref{app:vertices} Summing up the terms \eqref{eq:GSE}, we find
\begin{equation}\label{eq:1LoopPi}
 \Pi\ =\ \frac{\di}{8\kappa}\left[{F_{AC}}^D{F_{BD}}^C-2{T_{AI}}^J{T_{BJ}}^I\right]\int\frac{\dd^3 p}{(2\pi)^3}\,\dd^4\theta\, V^A(-p,\theta)\, p\, \CCP_{1/2}V^B(p,\theta)~.
\end{equation}
Note that the longitudinal part $\CCP_0$ does not appear in this expression as required by the Ward identity for the vector superfield propagator.

Using the result \eqref{eq:1LoopPi}, we can now derive the contribution to the anomalous dimension
of $\Phi^i$ coming from diagram \ref{fig:ZDiagrams}c). After some algebraic manipulations, we arrive
at
\begin{equation}\label{eq:Zc}
 \mbox{(c) :}~~~~\delta Z_i^{~j}\ =\ 
 -\frac{\log\Lambda}{48\pi^2\kappa^2}\big[2k_2+N_fk_3\big]\delta_i^{~j}~.
\end{equation}

Collecting all the results, \eqref{eq:Za}, \eqref{eq:Zb} and \eqref{eq:Zc}, we finally obtain 
\begin{equation}\label{eq:2LoopGammaR}
\begin{aligned}
 \gamma_i^{~j}\ &=\ \frac{1}{8\pi^2\kappa^2}
       \left\{\big[k_2+k_1^2+\tfrac{1}{12}(2k_2+N_fk_3)\big]\delta_i^{~j}\right.\\
   &\kern2.5cm
  +~8\kappa^2\left[R_{iklm}^{(1)}\big(-c_3R^{jklm}_{(1)}+2c_2 R^{jmlk}_{(1)}+2c_1R^{jmlk}_{(2)}\big)\right.\\
&\kern5cm\left.\left.+\ R_{iklm}^{(2)}\big(d\,R^{jklm}_{(2)}+2 R^{jmlk}_{(2)}+2c_1R^{jmlk}_{(1)}\big)
       \right]\right\}~
\end{aligned}
\end{equation}
for the anomalous dimension of $\Phi^i$.  Equation \eqref{eq:2LoopGammaR} may then be substituted into \eqref{eq:BetaFunction} to get
 the final expressions for the two-loop beta functions $\beta_{ijkl}^{(\ell)}$. 

As a check, let us consider $\CA=A_4$. In this case we have $d=4$ and $f_{abcd}=\varepsilon_{abcd}$. 
Then $k_1=0$, $k_2=-3$, $k_3=6$, $c_1=0$ and $c_2=c_3=-6$. We also take $N_f=4$ together with
$R_{ijkl}^{(1)}=\lambda \varepsilon_{ijkl}$ with some constant $\lambda$  and $R_{ijkl}^{(2)}=0$.
Using \eqref{eq:2LoopGammaR}, the beta functions \eqref{eq:BetaFunction} reduce to
\begin{equation}\label{eq:BLGBetaFunction}
 \beta_{ijkl}^{(1)}\ =\ -\tfrac{3}{4\pi^2\kappa^2}\big[1-(4!\kappa)^2|\lambda|^2\big]R^{(1)}_{ijkl}
 \eand \beta_{ijkl}^{(2)}\ =\ 0~,
\end{equation}
and this expression vanishes for either $\lambda=0$ (because $R^{(1)}_{ijkl}=\lambda \varepsilon_{ijkl}$) or $|\lambda|=\frac{1}{4!\kappa}$. The latter value of $\lambda$ is precisely the value for the original BLG model \eqref{eq:BLGmodel}. Furthermore, one might check that the phase of $\lambda$ does not flow (see also Section \ref{sec:DiscAndRes}). To characterize the fixed points, it is therefore sufficient to consider the modulus of $\lambda$. The value $|\lambda|=0$, the minimally coupled Chern-Simons matter theory, is thus a UV stable fixed point, while $|\lambda|=\frac{1}{4!\kappa}$, the BLG model, forms an IR stable fixed point.

\subsection{Two-loop renormalization in the Hermitian case}\label{eq:BetaHC}

Let us now discuss the Hermitian case with the action given by \eqref{eq:S0Complex}, \eqref{eq:S1Complex}, \eqref{eq:gaugefixingaction} and \eqref{eq:ghostaction}.
The calculation is essentially the same as in the
real case modulo some changes in the color/flavor structure of the diagrams due
to the two different types of matter that transform in opposite representations
of the gauge group. 

We introduce again
\begin{equation}
 \lbr X,Y\rbr\ =\ X^{ab} Y^{bc} f_{cabd}\ =:\ X^A Y^B G_{AB}~,\ewith G_{AB}\ =\ G_{BA}~
\end{equation}
and assume that $G_{AB}$ has an inverse. Due to the $ad$-invariance of
$\lbr\cdot,\cdot\rbr$, the structure constants $F_{ABC}:={F_{AB}}^D G_{CD}$ are totally
antisymmetric, as in the real case. Here, we have to use multi-indices of two types: $I=am$ and $\dot I=\,\!^a_\mdt$. Correspondingly, the chiral superfields
read as $\Phi^I$ and $\Phi_{\dot I}$ and their conjugates are $\Phib_I$ and $\Phib^{\dot I}$; in
writing this, we are implicitly using the metric $h_{ab}$ as we did in the real setting.
With these conventions, the propagators are essentially the same as those listed in
\eqref{eq:propagators}. The vertices are displayed in Appendix \ref{app:vertices} Everything else like regularization and power counting works, of course, as in the real setting. 

The beta functions for the two couplings $H_{mn\mdt\ndt}^{(\ell)}$ with $\ell=1,2$ are here given by
\begin{subequations}
\begin{equation}
\begin{aligned}
 \beta_{mn\mdt\ndt}^{(\ell)}\ &=\ 
      {(Z^{-1})_{mn\mdt\ndt}}^{m'n'\mdt'\ndt'}\gamma_{m'}^{~k}{Z^{(\ell)}_{kn'\mdt'\ndt'}}\,
     \!^{m''n''\mdt''\ndt''}H_{m''n''\mdt''\ndt''}^{(\ell)}\\
         &\kern1cm +\cdots+         
       {(Z^{-1})_{mn\mdt\ndt}}^{m'n'\mdt'\ndt'}\gamma_{\ndt'}^{~\dot k}
     {Z^{(\ell)}_{m'n'\mdt'\dot k}}\,\!^{m''n''\mdt''\ndt''}
       H_{m''n''\mdt''\ndt''}^{(\ell)}\\
       &\kern2cm+{((Z^{(\ell)})^{-1})_{mn\mdt\ndt}}\,\!^{m'n'\mdt'\ndt'}
 \frac{\dd {Z^{(\ell)}_{m'n'\mdt'\ndt'}}\,\!^{m''n''\mdt''\ndt''}}{\dd\log \mu}
H_{m''n''\mdt''\ndt''}^{(\ell)}~,
\end{aligned}
\end{equation}
where
\begin{equation}
\begin{aligned}
 \gamma_m^{~n}\ =\ 
                   \frac12 \frac{\dd(\log Z)_m^{~n}}{\dd\log \mu}\eand
\gamma_\mdt^{~\ndt}\ =\ 
                   \frac12 \frac{\dd(\log Z)_\mdt^{~\ndt}}{\dd\log \mu}
\end{aligned}
\end{equation}
\end{subequations}
denote the anomalous dimensions of the fields $\Phi^m$ and $\Phi^\mdt$ and 
${Z^{(\ell)}_{mn\mdt\ndt}}\,\!^{m'n'\mdt'\ndt'}$ 
is the renormalization of the quartic vertex $\ell$. As in the real
case, there is no renormalization of the vertices to two-loop order and we are therefore left with the wave function renormalizations
\begin{equation}\label{eq:BetaFunctionH}
 \beta_{mn\mdt\ndt}^{(\ell)}\ =\ \gamma_m^{~k}R_{kn\mdt\ndt}^{(\ell)}+
  \cdots+\gamma_\ndt^{~\dot k}R_{mn\mdt\dot k}^{(\ell)}~.
\end{equation}

Using the conventions introduced above, the diagrams in Fig.\ \ref{fig:ZDiagrams} yield the following
contributions to the wave function renormalization:
\begin{subequations}
\begin{eqnarray}
 \kern-1cm
 \mbox{(a) :}~~~~\delta Z_m^{~n}\! &=&\! -\frac{\log\Lambda}{4\pi^2\kappa^2}\big[k_2+k_1^2\big]\delta_m^{~n}\eand
 \delta Z_\mdt^{~\ndt}\ =\ -\frac{\log\Lambda}{4\pi^2\kappa^2}\big[k_2+
 k_1^2\big]\delta_\mdt^{~\ndt}~,\\
 \kern-1cm
 \mbox{(b) :}~~~~\delta Z_m^{~n}\! &=&\! 
   -\frac{\log\Lambda}{4\pi^2}\left[
     \big(H_{mk\mdt\ndt}^{(1)} H^{\mdt\ndt kn}_{(1)}-H_{mk\mdt\ndt}^{(1)} H^{\ndt\mdt kn}_{(1)}\big)
        c_2\cos^2\beta\right.\notag\\
     &&\kern1.5cm\left. +\ 
     \big(H_{mk\mdt\ndt}^{(1)} H^{\mdt\ndt kn}_{(1)}+H_{mk\mdt\ndt}^{(1)} H^{\ndt\mdt kn}_{(1)}\big)
        c'_2\sin^2\beta\right.\notag\\
     &&\kern1.5cm\left. +\ 
      \big(H_{mk\mdt\ndt}^{(1)}H^{\mdt\ndt kn}_{(2)}+H_{mk\mdt\ndt}^{(2)} H^{\mdt\ndt kn}_{(1)}\big)
        \big(c_1\cos\beta+\di c'_1\sin\beta\big)\right.\notag\\
     &&\kern1.5cm \left. -\ 
     \big(H_{mk\mdt\ndt}^{(1)}H^{\ndt\mdt kn}_{(2)}+H_{mk\mdt\ndt}^{(2)} H^{\ndt\mdt kn}_{(1)}\big)
        \big(c_1\cos\beta-\di c'_1\sin\beta\big)\right.\notag\\
      &&\kern1.5cm\left.+\ 
     \big(H_{mk\mdt\ndt}^{(2)}H^{\mdt\ndt kn}_{(2)}+d\,H_{mk\mdt\ndt}^{(2)} H^{\ndt\mdt kn}_{(2)}\big)\right],\label{Zb1}\\
\delta Z_\mdt^{~\ndt}\! &=&\! 
   -\frac{\log\Lambda}{4\pi^2}\left[
     \big(H^{\ndt\dot kmn}_{(1)} H_{mn\dot k\mdt}^{(1)}-H^{\dot k\ndt mn}_{(1)} H_{mn \dot k\mdt}^{(1)}\big)
        c_2\cos^2\beta\right.\notag\\
     &&\kern1.5cm\left. +\ 
     \big(H^{\ndt\dot kmn}_{(1)} H_{mn\dot k\mdt}^{(1)}+H^{\dot k\ndt mn}_{(1)} H_{mn \dot k\mdt}^{(1)}\big)
        c'_2\sin^2\beta\right.\notag\\
     &&\kern1.5cm\left. +\ 
      \big(H^{\ndt\dot kmn}_{(1)}H_{mn\dot k\mdt}^{(2)}+H^{\ndt\dot k mn}_{(2)} H_{mn\dot k\mdt}^{(1)}\big)
        \big(c_1\cos\beta+\di c'_1\sin\beta\big)\right.\notag\\
     &&\kern1.5cm \left. -\ 
     \big(H^{\dot k\ndt mn}_{(1)}H_{mn\dot k\mdt}^{(2)}+H^{\dot k\ndt mn}_{(2)} H_{nm\dot k\mdt}^{(1)}\big)
        \big(c_1\cos\beta-\di c'_1\sin\beta\big)\right.\notag\\
      &&\kern1.5cm\left.+\ 
     \big(H^{\ndt\dot kmn}_{(2)}H_{mn\dot k\mdt}^{(1)}+d\,H^{\dot k\ndt mn}_{(2)} 
      H_{mn\dot k\mdt}^{(2)}\big)\right],\label{Zb2}\\
\kern-1cm
\mbox{(c) :}~~~~\delta Z_m^{~n}\! &=&\! -\frac{\log\Lambda}{48\pi^2\kappa^2}
  \big[2k_2+N_fk_3\big]\delta_m^{~n}\eand
 \delta Z_\mdt^{~\ndt}\ =\ -\frac{\log\Lambda}{48\pi^2\kappa^2}
  \big[2k_2+N_fk_3\big]\delta_\mdt^{~\ndt}~,
\end{eqnarray}
where
\begin{equation}
\begin{aligned}
 (\CCO_1)_I^{~J}\ &:=\ G^{AB}{T_{AI}}^K{T_{BK}}^L G^{CD}{T_{CL}}^M{T_{DM}}^J\ =\ k^2_1\delta_I^{~J}~,\\
 (\tilde \CCO_1)_{\dot I}^{~\dot J}\ &:=\ G^{AB}{T_{A\dot I}}^{\dot K}{T_{B\dot K}}^{\dot L}
 G^{CD}{T_{C\dot L}}^{\dot M}{T_{D\dot M}}^{\dot J}\ =\ k^2_1\delta_{\dot I}^{~\dot J}~,\\
 (\CCO_2)_I^{~J}\ &:=\ \tfrac14 G^{AC}G^{BD} {F_{AB}}^E {F_{CD}}^F {T_{EI}}^{~K} {T_{FK}}^{~J}
  \ =\ k_2\delta_I^{~J}~,\\
 (\tilde \CCO_2)_{\dot I}^{~\dot J}\ &:=\ \tfrac14 G^{AC}G^{BD} {F_{AB}}^E {F_{CD}}^F {T_{E\dot I}}^{~\dot K} 
  {T_{F\dot K}}^{~\dot J} \ =\ k_2\delta_{\dot I}^{~\dot J}~,\\
 (\CCO_3)_I^{~J}\ &:=\ \tfrac{2}{N_f}{T_{AK}}^L {T_{BL}}^K G^{AC} G^{BD}{T_{CI}}^M {T_{DM}}^J
                  =\  k_3 \delta_I^{~J}~,\\
 (\tilde \CCO_3)_{\dot I}^{~\dot J}\ &:=\ \tfrac{2}{N_f}{T_{A\dot K}}^{\dot L} {T_{B\dot L}}^{\dot K} 
 G^{AC} G^{BD}{T_{C\dot I}}^{\dot M} {T_{D\dot M}}^{\dot J}
                  =\ k_3 \delta_{\dot I}^{~\dot J}~,
\end{aligned}
\end{equation}
and 
\begin{equation}
 {f_{ac}}^{cb}\ =\ c_1\delta_a^{~b}~,~~~f_{acde}f^{edcb}\ =\ -c_2{\delta_a}^b~,~~~
 {d_{ac}}^{cb}\ =\ c'_1\delta_a^{~b}~,~~~
 d_{acde}d^{edcb}\ =\ -c'_2{\delta_a}^b~
\end{equation}
\end{subequations}
with $d=\operatorname{dim}\CA$. For the explicit values of the Casimirs $k_i$, $c_i$ and $c'_i$ in the matrix representation $M_{\mathrm{I}_\alpha}^H(N)$, we refer to Appendix \ref{app:UsefulFormulae}

Altogether, we obtain the following anomalous dimensions:
\begin{subequations}\label{eq:2LoopGammaH}
 \begin{eqnarray}
  \gamma_m^{~n}\! &=&\! \frac{1}{8\pi^2\kappa^2}\bigg\{\big[k_2+k_1^2+\tfrac{1}{12}(2k_2
   +N_fk_3)\big]\delta_m^{~n}\notag\\
   &&\kern1cm+~\kappa^2\left[
     \big(H_{mk\mdt\ndt}^{(1)} H^{\mdt\ndt kn}_{(1)}-H_{mk\mdt\ndt}^{(1)} H^{\ndt\mdt kn}_{(1)}\big)
        c_2\cos^2\beta\right.\notag\\
     &&\kern1.5cm\left. +\ 
     \big(H_{mk\mdt\ndt}^{(1)} H^{\mdt\ndt kn}_{(1)}+H_{mk\mdt\ndt}^{(1)} H^{\ndt\mdt kn}_{(1)}\big)
        c'_2\sin^2\beta\right.\notag\\
     &&\kern1.5cm\left. +\ 
      \big(H_{mk\mdt\ndt}^{(1)}H^{\mdt\ndt kn}_{(2)}+H_{mk\mdt\ndt}^{(2)} H^{\mdt\ndt kn}_{(1)}\big)
        \big(c_1\cos\beta+\di c'_1\sin\beta\big)\right.\notag\\
     &&\kern1.5cm \left. -\ 
     \big(H_{mk\mdt\ndt}^{(1)}H^{\ndt\mdt kn}_{(2)}+H_{mk\mdt\ndt}^{(2)} H^{\ndt\mdt kn}_{(1)}\big)
        \big(c_1\cos\beta-\di c'_1\sin\beta\big)\right.\notag\\
      &&\kern1.5cm\left.+\ 
     \big(H_{mk\mdt\ndt}^{(2)}H^{\mdt\ndt kn}_{(2)}+d\,H_{mk\mdt\ndt}^{(2)} H^{\ndt\mdt kn}_{(2)}\big)\right]\bigg\}~,\\
\gamma_{\dot m}^{~\dot n}\! &=&\! \frac{1}{8\pi^2\kappa^2}\bigg\{\big[k_2+k_1^2+
 \tfrac{1}{12}(2 k_2
   +N_fk_3)\big]\delta_m^{~n}\notag\\
   &&\kern1cm+~\kappa^2\left[
     \big(H^{\ndt\dot kmn}_{(1)} H_{mn\dot k\mdt}^{(1)}-H^{\dot k\ndt mn}_{(1)} H_{mn \dot k\mdt}^{(1)}\big)
        c_2\cos^2\beta\right.\notag\\
     &&\kern1.5cm\left. +\ 
     \big(H^{\ndt\dot kmn}_{(1)} H_{mn\dot k\mdt}^{(1)}+H^{\dot k\ndt mn}_{(1)} H_{mn \dot k\mdt}^{(1)}\big)
        c'_2\sin^2\beta\right.\notag\\
     &&\kern1.5cm\left. +\ 
      \big(H^{\ndt\dot kmn}_{(1)}H_{mn\dot k\mdt}^{(2)}+H^{\ndt\dot k mn}_{(2)} H_{mn\dot k\mdt}^{(1)}\big)
        \big(c_1\cos\beta+\di c'_1\sin\beta\big)\right.\notag\\
     &&\kern1.5cm \left. -\ 
     \big(H^{\dot k\ndt mn}_{(1)}H_{mn\dot k\mdt}^{(2)}+H^{\dot k\ndt mn}_{(2)} H_{nm\dot k\mdt}^{(1)}\big)
        \big(c_1\cos\beta-\di c'_1\sin\beta\big)\right.\notag\\
      &&\kern1.5cm\left.+\ 
     \big(H^{\ndt\dot kmn}_{(2)}H_{mn\dot k\mdt}^{(1)}+d\,H^{\dot k\ndt mn}_{(2)} 
      H_{mn\dot k\mdt}^{(2)}\big)\right]\bigg\}~.
 \end{eqnarray}
\end{subequations}
These expressions may be substituted into
\eqref{eq:BetaFunctionH} to arrive at the final result for the beta functions.

As a check, let us consider the ABJM model. In that case we have,
$\beta=0$, $N_f=4$, $H_{mn\mdt\ndt}^{(1)}=\lambda\varepsilon_{mn}\varepsilon_{\mdt\ndt}$
for some constant $\lambda$ and $H^{(2)}_{mn\mdt\ndt}=0$. Furthermore, we choose $M_{\mathrm{I}_{\alpha=1}}^H(N)$
and hence $k_1=0$, $k_2=1-N^2 $, $k_3=-2+2N^2$, $c_1=0$ and $c_2=2-2N^2$. Therefore, we find
\begin{equation}
\begin{aligned}
 \gamma_m^{~n}\ &=\ \frac{1}{16\pi^2\kappa^2}(1-N^2)\big[1-(4\kappa)^2|\lambda|^2\big]\delta_m^{~n}~,\\ 
 \gamma_\mdt^{~\ndt}\ &=\ \frac{1}{16\pi^2\kappa^2}(1-N^2)\big[1-(4\kappa)^2|\lambda|^2\big]\delta_\mdt^{~\ndt}~,
\end{aligned}
\end{equation}
and thus, we recover precisely the value $|\lambda|=\frac{1}{4\kappa}$ for the ABJM model;
see equations \eqref{eq:ABJMmodel}. For $N=2$, this of course agrees with the result
\eqref{eq:BLGBetaFunction} as for this particular value of $N$, the ABJM model coincides with the BLG model. As in the real case, the phase of $\lambda$ does not flow (see also Section \ref{sec:DiscAndRes}) and so we can restrict ourselves to the modulus $|\lambda|$. Therefore, the conformal fixed point corresponding to the ABJM model forms an IR fixed point, just like in the case of the BLG model.

\subsection{Discussion of the results}\label{sec:DiscAndRes}

The above expressions for the anomalous dimensions and the resulting expressions for the beta functions certainly allow for many conformal fixed points depending on the particular choices of the superpotential couplings and of the underlying 3-algebra structure. For this reason, we shall merely discuss two examples. We hope to report on a more thorough analysis elsewhere. In our subsequent discussion, we assume that $N_f=4$.

\bigskip
\noindent{\it \underline{Real 3-algebras:}}
\bigskip

\noindent
Let us consider $\CA=A_4$. We recall that in this case the Casimirs are given by
\begin{equation}
 k_1\ =\ 0~,~~~k_2\ =\ -3~,~~~k_3\ =\ 6~,~~~c_1\ =\ 0~,~~~c_2\ =\ -6~,~~~c_3\ =\ -6~.
\end{equation}
Furthermore, we take
\begin{equation}\label{rels}
 R_{ijkl}^{(1)}\ =\ \frac{\lambda_1}{\kappa}\,\varepsilon_{ijkl}\eand
 R_{ijkl}^{(2)}\ =\ \frac{\lambda_2}{\kappa}\, \delta_{ij}\delta_{kl}~,
\end{equation}
with $\lambda_\ell=r_\ell \de^{\di\varphi_\ell}$.
Plugging these values into the expression \eqref{eq:2LoopGammaR} for the two-loop
anomalous dimension, one finds that the corresponding beta functions are given by 
\begin{equation}\label{bet}
\beta^{(\ell)}_{ijkl}\ =\ \frac{f(r_1,r_2)}{\kappa^2} R_{ijkl}^{(\ell)}~,\ewith
f(r_1,r_2)\ :=\ -\frac{3}{4\pi^{2}}\left[1-96\big(6r_1^{2} + r_2^{2}\big)\right]~.
\end{equation}
The zero-locus $f(r_1,r_2)=0$ defines an ellipse in $\FR^2$,
\begin{equation}
 r_1\ =\ \frac{1}{24}\cos t\eand
 r_2\ =\ \frac{1}{4\sqrt{6}}\sin t \efor t\ \in\ [0,2\pi)~.
\end{equation}
We thus obtain a one-parameter family of marginal multi-trace deformations (i.e.\ double-trace in superfields 
and double- and triple-trace in components) of the BLG model, the latter corresponding to $t=0$. 

\vspace*{2cm}

\begin{figure}[h]
\hspace{3.5cm}
\begin{picture}(240,150)
\psfrag{0.00}{\kern-2pt 0.0}
\psfrag{0.0}{\kern-2pt 0.0}
\psfrag{0.05}{\kern-7pt 0.05}
\psfrag{0.10}{\kern-3pt 0.1}
\psfrag{0.1}{\kern-3pt 0.1}
\psfrag{0.2}{\kern-2pt 0.2}
\psfrag{0.15}{\kern-3pt 0.15}
\includegraphics[width=90mm]{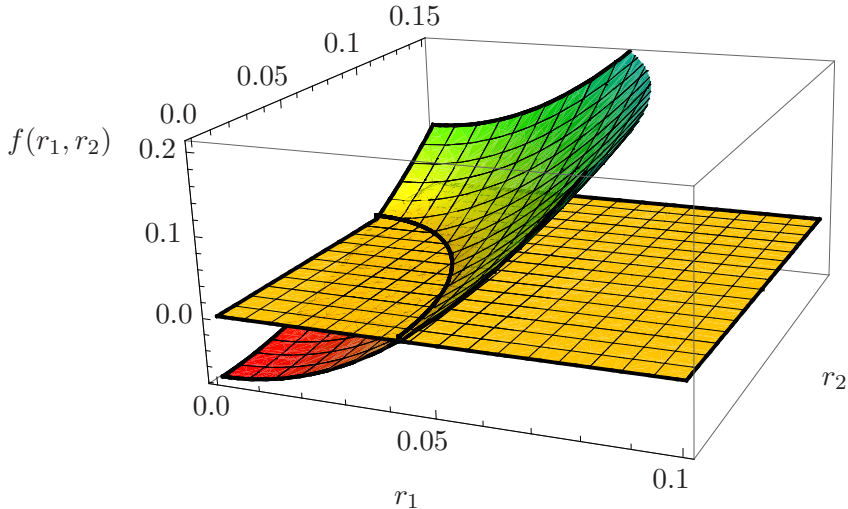}
\put(0.0,60.0){\makebox(0,0)[c]{$r_2$}}
\put(-290.0,150.0){\makebox(0,0)[c]{$f(r_1,r_2)$}}
\put(-160.0,15.0){\makebox(0,0)[c]{$r_1$}}
\end{picture}
\vspace*{-5pt}
\caption{The function $f(r_1,r_2)$ capturing the beta functions of single- and multi-trace deformations.}\label{fig2}
\end{figure}
\vspace*{10pt}

Furthermore, \eqref{bet} implies the following equations for the running couplings $\tilde\lambda_\ell=\tilde r_\ell\,\de^{\di \tilde\varphi_\ell}$:
\begin{equation}
 \dot{\tilde{r}}_\ell\ =\ \frac{\tilde{r}_\ell }{\kappa^2}f(\tilde{r}_1,\tilde{r}_2)\eand \tilde{r}_\ell\dot{\tilde{\varphi}}_\ell\ =\ 0~,\ewith \tilde{\lambda}_\ell(\mu;\lambda_\ell)\ =\ \lambda_\ell~,
\end{equation}
where dot means a total derivative with respect to $\log (p/\mu)$. Hence, the phases 
$\tilde{\varphi}_\ell$ do not flow. 
To get a more intuitive picture of the situation, we plotted the function $f(r_1,r_2)$ in Fig.\ \ref{fig2} for $r_1,r_2>0$. 
{}From the figure it is then clear that every point on the fixed point locus of the beta functions corresponds to an IR fixed point of the renormalization group, as the derivative of the function $f(r_1,r_2)$ in the direction of the outward normal of the curve is positive. Notice further that the minimally coupled Chern-Simons matter theory, $r_\ell=0$, is a UV fixed point. Thus, by turning on the above deformation at $r_\ell=0$, the theory flows to one of the points on the curve $f(r_1,r_2)=0$ in the IR.

\bigskip
\noindent{\it \underline{Hermitian 3-algebras:}}
\bigskip

\noindent
Let us now perform a similar analysis in the Hermitian setting. We take $\CA=M^H_{\mathrm{I}_{\alpha=1}}(N)$.
In this case, we know that
\begin{equation}
\begin{aligned}
 &k_1\ =\ 0~,~~~k_2\ =\ 1-N^2~,~~~k_3\ =\ -2(1-N^2)~,~~~c_1\ =\ 0~,~~~c_2\ =\ 2(1-N^2)~,\\
 &\kern4.0cm c'_1\ =\ -2N ~,~~~c'_2\ =\ -2(1+N^2)~
\end{aligned}
\end{equation}
and $d=N^2$.
Let us focus on superpotential couplings of the form
\begin{equation}
\begin{aligned}
 H_{mn\mdt\ndt}^{(1)}\ &=\ \frac{\lambda_1}{\kappa}\big[\varepsilon_{mn}\varepsilon_{\mdt\ndt}
                           +\rho(\delta_{(mn\mdt\ndt),(1,2,2,1)}+\delta_{(mn\mdt\ndt),(2,1,1,2)})\big]~,\\
 H_{mn\mdt\ndt}^{(2)}\ &=\ \frac{\lambda_2}{\kappa}\, \delta_{m\mdt}\delta_{n\ndt}~.
\end{aligned}
\end{equation}
Note that $\lambda_2$ controls the multi-trace deformations. 
Substituting these expressions into \eqref{eq:2LoopGammaH} for the two-loop anomalous dimension, we find after some algebraic manipulations that the beta functions \eqref{eq:BetaFunctionH} vanish if
($\lambda_\ell=r_\ell\,\de^{\di \varphi_\ell}$ and $\rho=r_3\,\de^{\di\varphi_3}$)
\begin{subequations}
\begin{equation}\label{vanish1}
 a\,r_1^2(r_3^2-4r_3\cos\varphi_3+4)+b\,r_1^2r_3^2+c\,
 r_2^2+d\,r_1r_2r_3\sin(\varphi_1-\varphi_2+\varphi_3)\ =\ 1~,
\end{equation}
where
\begin{equation}
 a\ :=\ 4\cos^2\beta~,~~~
 b\ :=\ \frac{2N^2+2}{N^2-1}\sin^2\beta~,~~~
 c\ :=\ \frac{4N^2+2}{N^2-1}~,~~~
 d\ :=\ \frac{8N}{N^2-1}\sin\beta~.
\end{equation}
\end{subequations}
For $\rho=-2$ and $\lambda_2=0$, we find the $\beta$-deformed ABJM model that was discussed in \cite{Imeroni:2008cr} by studying the gravitational dual of the theory while for $\rho=0$ (implying
$\beta=0$ without loss of generality) and $\lambda_2\neq0$, we obtain a marginal multi-trace deformation of
the ABJM model.

\section{Conclusions and outlook}\label{sec:conclusions}

In summary, we have described marginal deformations of Chern-Simons matter theories that are based on real and Hermitian 3-algebras. In particular, we wrote down the most general superpotentials consisting of single- and multi-trace terms that are i) conformally invariant at the classical level, ii) compatible with $\CN=2$ supersymmetry
and iii) supergauge invariant. For these superpotential terms, we computed the two-loop beta functions using $\CN=2$ supergraph techniques. As familiar from four dimensional SYM theories, supergraphs turned out to be a powerful tool also in the case of supersymmetric Chern-Simons matter theories:
The calculation of the two-loop beta functions boiled down to the computation of the three Feynman supergraphs displayed in Fig.\ \ref{fig:ZDiagrams}. We expressed our results concisely in terms of certain ``Casimir invariants'' of the underlying 3-algebra and
its associated Lie algebra. Using our expressions for the beta functions, we confirmed conformality of both the BLG and ABJM models. In addition, we discussed $\beta$-deformations of the 
ABJM model and certain marginal multi-trace deformations of both the BLG
and ABJM models, explicitly. We mostly focused
on the 3-algebras $M^R_{\rm{III}_{\alpha,\beta}}(N)$ and $M^H_{\rm{I}_\alpha}(N)$, but
a similar analysis can easily be carried out for other 3-algebras.

Even though real and Hermitian 3-algebras already allow for classes of  marginal deformations, we found that not all deformations, and in particular not the $\beta$-deformations of \cite{Imeroni:2008cr}, are captured by 3-brackets. Instead, one has to introduce an associated 3-product, i.e.\ a triple product that transforms covariantly under gauge transformations. This is in the same spirit
to what happens in four-dimensional SYM theory, where one replaces the Lie bracket by some deformed bracket. To discuss $\beta$-deformations of the ABJM model, for instance, we were led to introduce the $\beta$-3-bracket \eqref{eq:BetaCommutator}, which is
just a special instance of an associated 3-product. As far as $\beta$-deformations are concerned, we mainly focused on the Hermitian case. Here, we obtained an independent confirmation of the deformations studied in \cite{Imeroni:2008cr}. Note, however, that more general deformations than the $\beta$-deformations we focused on can in principle be discussed in both the real and Hermitian cases using associated 3-products.

The most interesting open question is certainly to what extent our
deformations are exactly marginal, or at least, to all orders in perturbation theory. Because of the many simplifications which arise, e.g., due to Lemma 3 of \cite{Gates:1991qn}, one might be able to make precise statements using our superfield formulation. Otherwise, it might be necessary to switch to a different description as, for example, light-cone superspace as done in \cite{Ananth:2006ac} for $\beta$-deformations of $\CN=4$ SYM theory.

Another point is certainly to study the 't Hooft limit of our deformed theories\footnote{after
appropriate rescalings by factors of `$N$', see e.g.~\cite{Witten:2001ua}} and identify all geometries which form their gravitational duals, extending the work of \cite{Imeroni:2008cr}. Vice versa, one could reformulate some of the deformations considered in \cite{Imeroni:2008cr} in terms of 3-algebra language to gain more insight into the 3-algebra structures involved.

Finally, it would be interesting to extend the analysis of  \cite{Minahan:2008hf,Bak:2008cp} to our deformed BLG-type models and to study a possible correspondence of the dilatation operator in these models to the Hamiltonian of an integrable spin chain, using superspace and 3-algebra language. This is possible, because the operators considered in \cite{Minahan:2008hf} can easily be formulated in terms of the associated 3-products
introduced in this work.

\vspace*{.5cm}
\noindent
{\bf Acknowledgements.}
We are very grateful to Martin Ammon, Sergey Cherkis, Stefano Kovacs and Riccardo Ricci for discussions, questions and suggestions.
N.A.\ was supported by the Dutch Foundation for Fundamental Research on Matter (FOM).
C.S.\ was supported by an IRCSET Postdoctoral Fellowship.
M.W.\ was supported by an STFC Postdoctoral Fellowship and by a Senior Research
Fellowship at the Wolfson College, Cambridge, U.K.

\appendices

\subsection{Casimirs for matrix representations}\label{app:UsefulFormulae}

In this appendix, we discuss the Casimirs $k_i$ and $c_i$ that appear throughout this
work for the different matrix representations. 

\bigskip
\noindent{\it \underline{Casimirs $c_i$ and $k_i$ for the real 3-algebra  $M_{\rm III_{\alpha,\beta}}^R(N)$:}}
\bigskip

\noindent 
The underlying vector space for this real 3-algebras has dimension $N^2$ and one easily finds a basis with elements satisfying the following relations
\begin{equation}
 \tr (\tau_a^T \tau_b)\ =\ \delta_{ab}\ =:\ 
  h_{ab}\eand(\tau_a)_{ij}(\tau_a)_{kl}\ =\ \delta_{ik}\delta_{jl}~.
\end{equation}
With the above formul\ae{}, the three Casimirs $c_1,c_2$ and $c_3$ defined by 
\begin{equation}
 {f_{ac}}^{cb}\ =\ c_1{\delta_a}^b~,~~~
 f_{acde}f^{bedc}\ =\ c_2{\delta_a}^b\eand
 f_{acde}f^{bcde}\ =\ -c_3\delta_a^{~b}
\end{equation}
 can be  computed straightforwardly and we obtain
\begin{equation}
\begin{aligned}
 &c_1\ =\ (N-1)(\alpha-\beta)~,~~~
 c_2\ =\ (N-1)(\alpha^2-2(N-1)\alpha\beta+\beta^2)~,\\
 &\kern3cm c_3\ =\ -2N(N-1)(\alpha^2+\beta^2)~.
\end{aligned}
\end{equation}

The Casimirs $k_i$ can similarly be calculated by using identities for the appearing generators of $\frg_\CA\cong\ao(N)\oplus\ao(N)$ together with formula \eqref{signature-real} for the bilinear form  $\lbr\cdot,\cdot\rbr$ on $\frg_\CA$. We find here that
\begin{equation}
 k_1\ =\ \tfrac{1}{\sqrt{2}}(\alpha^3+\beta^3)~,~~~k_2\ =\ -\tfrac{1}{4}(\alpha^3+\beta^3)~,~~~k_3\ =\ -\tfrac{1}{2}N(\alpha^6+\beta^6)~.
\end{equation}
Note that the algebra $A_4$ is a sub-3-algebra of the 3-algebra $M^R_{{\rm III}_{1,-1}}(4)$. In this case, one can compute the Casimirs directly from the structure constants and the fact that $\frg_\CA=\asu(2)\oplus\asu(2)$ and we obtain
\begin{equation}\label{A4Casimirs}
 c_1\ =\ 0~,~~~c_2\ =\ -6~,~~~c_3\ =\ -6~,~~~k_1\ =\ 0~,~~~k_2\ =\ -3~,~~~k_3\ =\ 6~.
\end{equation}

Analogously, one constructs the Casimirs for the other real 3-algebras $M^R_{\rm I_\alpha}(N)$, $M^R_{\rm II_\alpha}(N)$ and $M^R_{\rm IV_{\alpha,\beta}}(N)$, but we refrain from going into more detail at this point.

\bigskip
\noindent{\it \underline{Casimirs $c_i$ and $k_i$ for the Hermitian 3-algebra $M^H_{\rm I_\alpha}(N)$:}}
\bigskip

\noindent  
The underlying vector space here is spanned by generators of $\sU(N)$ in the fundamental representation. For simplicity, we fix $\alpha=1$, as we did throughout most of the paper. As basis $\tau_a$, we have $N^2$ anti-Hermitian $N\times N$-matrices and we choose them such that we have the following identities:
\begin{equation}
 \tr (\tau_a^\dagger \tau_b)\ =\ 
  \delta_{ab}\ =:\ h_{ab}~,~~~h^{ab}\ =\ \delta^{ab}\eand(\tau_a)_{ij}(\tau_a)_{kl}\ =\ -\delta_{il}\delta_{jk}~.
\end{equation}
{}From these, one obtains for the Casimirs $c_1,c_2$ and $c_3$, which are defined by 
\begin{equation}
 {f_{ac}}^{cb}\ =\ c_1{\delta_a}^b~,~~~
 f_{acde}f^{edcb}\ =\ -c_2{\delta_a}^b\eand
 f_{acde}f^{bcde}\ =\ -c_3\delta_a^{~b}~,
\end{equation}
the following expressions:
\begin{equation}\label{HCasimirs1}
 c_1\ =\ 0~,~~~c_2\ =\ 2(1-N^2)\eand c_3\ =\ 2(1-N^2)~.
\end{equation}
In addition, we have
\begin{equation}\label{HCasimirs2}
 k_1\ =\ 0~,~~~k_2\ =\ 1-N^2\eand k_3\ =\ -2(1-N^2)~.
\end{equation}
Recall that $M^H_{\rm I_{\alpha=1}}(2)=A_4$, and the above formul\ae{} \eqref{HCasimirs1} and \eqref{HCasimirs2} reproduce indeed \eqref{A4Casimirs} for $N=2$.

\bigskip
\noindent{\it \underline{Remarks on the bracket $[A,B;C]_\beta$:}}
\bigskip

\noindent 
Recall the form of the $\beta$-3-bracket
\begin{equation}
 [\tau_a,\tau_b;\tau_c]_\beta\ =\ {g_{abc}}^d\tau_d\ewith
 {g_{abc}}^d\ =\ \cos\beta {f_{abc}}^d+\di\sin\beta\, {d_{abc}}^d~.
\end{equation}
Therefore, apart from the Casimirs $c_i$ we also have the $c'_i$ 
\begin{equation}
 {d_{ac}}^{cb}\ =\ c'_1{\delta_a}^b\eand
 d_{acde}d^{edcb}\ =\ -c'_2{\delta_a}^b~.
\end{equation}
Explicitly, we obtain the following values:
\begin{equation}
 c'_1\ =\ -2 N\eand c_2'\ =\ -2(1+N^2)~.
\end{equation}

\subsection{Component form of the actions}\label{app:CFofA}

In this appendix we give the component form of the superspace actions in WZ gauge. 

\bigskip
\noindent{\it \underline{Component action in the real case:}}
\bigskip

\noindent 
In terms of the component fields \eqref{eq:WZGaugeComponents}, the action \eqref{eq:S0Real} reads as
\begin{equation}\label{S_{0}R}
\begin{aligned}
 S^R_0\ =\ &\int \dd^3 x\, 
   \bigg[\eps^{\mu\nu\lambda}\lbr A_\mu,\dpar_\nu A_\lambda+
      \tfrac{1}{3\sqrt{\kappa}}\lsb A_\nu,A_\lambda\rsb\rbr-
       \di \lbr\bl_\alpha,\lambda^\alpha\rbr-
       \di \lbr\lambda_\alpha,\bl^\alpha\rbr-
           \lbr D,\sigma\rbr-\lbr \sigma,D\rbr\\
      &~~~~~~+(\bar{F}_i,F^i)-
             (\nabla_\mu \bar{\phi}_i,\nabla^\mu \phi^i)-
            \di(\bar{\psi}_i^\alpha,\nabla_{\alpha\beta}\psi^{i\beta})-
            \tfrac{\di}{\sqrt{\kappa}} (\bar{\phi}_i,D(\phi^i))+
           \sqrt{\tfrac{2}{\kappa}} (\bar{\phi}_i,\lambda^\alpha(\psi^i_\alpha))\\
&\kern1cm +\sqrt{\tfrac{2}{\kappa}}(\bl_\alpha(\bar{\psi}_i^\alpha),\phi^i)+
          \tfrac{1}{\kappa}(\bar{\phi}_i,\sigma^2(\phi^i))+
            \tfrac{1}{\sqrt{\kappa}}(\bar{\psi}_{i\alpha},\sigma(\psi^{i\alpha}))\bigg]~,
\end{aligned}
\end{equation}
where $\nabla_{\alpha\beta}:=\sigma^\mu_{\alpha\beta}\nabla_\mu$. Upon performing the integrals over the fermionic directions, the component form of the superpotential term \eqref{eq:S1Real} is given by 
\begin{equation}\label{superterms}
\begin{aligned}
 S_1^R\ &=\ -2\int \dd^3x~\bigg\{
R_{ijkl}^{(1)}\left[\big(\phi^l,[\psi^{i\alpha },\psi_\alpha^j,\phi^k]+2[\phi^i,\psi^{j\alpha},\psi_\alpha^k]\big)-2([\phi^i,\phi^j,\phi^k],F^l)\right]\\
       &\kern2.6cm
 +R_{ijkl}^{(2)}\left[(\psi^{i\alpha },\psi_\alpha^j)(\phi^k,\phi^l)+
  2(\psi^{i\alpha},\phi^j)(\psi^k_\alpha,\phi^l)-2(F^i,\phi^j)(\phi^k,\phi^l)\right]
\bigg\}\\
     &\kern3.6cm+\mathrm{c.c.}~.
\end{aligned} 
\end{equation}
The next step is to eliminate the auxiliary fields. After varying $S^R=S_0^R+S_1^R$, 
we find the following (algebraic) equations
of motion for $F^i$, $\bar F_i$, $D$, $\sigma$, $\lambda$ and $\bar\lambda$:
\begin{equation}\label{eq:AuxEqReal}
 \begin{aligned}
    F^i\ &=\ -4R^{ijkl}_{(1)}[\bphi_l,\bphi_k,\bphi_j]-4R^{ijkl}_{(2)}(\bar\phi_l,\bar\phi_k)\bar\phi_j~,\\
   \bar F_i\ &=\ -4R_{ijkl}^{(1)}[\phi^l,\phi^k,\phi^j]-4R_{ijkl}^{(2)}(\phi^l,\phi^k)\phi^j~,\\
   D(A)\ &=\ \tfrac{1}{2\kappa}\big[[\phi^i,\sigma(\bphi_i),A]-
       [\sigma(\phi^i),\bphi_i,A]\big]
      +\tfrac{1}{2\sqrt{\kappa}}[\psi^{i\alpha},\bpsi_{i\alpha},A]~,\\
   \sigma(A)\ &=\ -\tfrac{\di}{2\sqrt{\kappa}}[\phi^i,\bphi_i,A]~,\\
   \lambda_\alpha(A)\ &=\ -\tfrac{\di}{\sqrt{2\kappa}}[\bpsi_{i\alpha},\phi^i,A]\eand
   \bar\lambda_\alpha(A)\ =\ \tfrac{\di}{\sqrt{2\kappa}}[\psi^i_{\alpha},\bphi_i,A] ~,
 \end{aligned}
\end{equation}
and hence $S^R=\int\dd^3x\, \CL^R $ with
\begin{equation}\label{eq:S01RComp}
\begin{aligned}
\CL^R\ &=\ \eps^{\mu\nu\lambda}\lbr A_\mu,\dpar_\nu
           A_\lambda+\tfrac{1}{3\sqrt{\kappa}}\lsb A_\nu,A_\lambda\rsb\rbr-\big|\nabla^\mu\phi^i\big|^2-
            \di\big(\bar{\psi}_i^\alpha,\nabla_{\alpha\beta} \psi^{i\beta}\big)\\[3pt]
&\kern.5cm+\tfrac{1}{4\kappa^2}\big([\phi^i,\bphi_i,\bphi_k],[\phi^j,\bphi_j,\phi^k]\big)+
\tfrac{\di}{2\kappa}\left(\bpsi_j^{\alpha},[\phi^i,\bphi_i,\psi^j_\alpha]\right)+\tfrac{\di}{\kappa}\left([\bpsi^\alpha_j,\phi^j,\bphi_i],\psi^i_\alpha\right)\\[3pt]
&\kern.5cm-2 R_{ijkl}^{(1)}\big(\phi^l,[\psi^{i\alpha},\psi_\alpha^j,\phi^k]+2[\phi^i,
  \psi^{j\alpha},\psi_\alpha^k]\big)\\[3pt]
&\kern.5cm-2R^{ijkl}_{(1)}\big(\bphi_l,[\bpsi_{i\alpha},\bpsi^\alpha_j,\bphi_k]+2[\bphi_i,\bpsi_{j\alpha },\bpsi^\alpha_k]\big)\\[3pt]
&\kern.5cm-2R_{ijkl}^{(2)}\Big[(\psi^{i\alpha },\psi_\alpha^j)(\phi^k,\phi^l)+
  2(\psi^{i\alpha},\phi^j)(\psi^k_\alpha,\phi^l)\Big]\\[3pt]
&\kern.5cm-2R^{ijkl}_{(2)}\Big[(\bar\psi_{i\alpha },\bar\psi^\alpha_j)(\bphi_k,\bphi_l)+
  2(\bar\psi_{i\alpha},\bphi_j)(\bar\psi_k^\alpha,\bphi_l)\Big]\\[3pt]
&\kern.5cm-16\left|R_{ijkl}^{(1)}[\phi^l,\phi^k,\phi^j]+R_{ijkl}^{(2)}(\phi^l,\phi^k)\phi^j \right|^2~,
 \end{aligned}
\end{equation}
where $|A|^2:=(\bar A,A)$ for any $A\in\CA$. 

For the reader's convenience, we finally extract the multi-trace terms explicitly:
\begin{equation}
\begin{aligned}
 \CL_{mult}^R\ &=\ -2R_{ijkl}^{(2)}\Big[(\psi^{i\alpha },\psi_\alpha^j)(\phi^k,\phi^l)+
  2(\psi^{i\alpha},\phi^j)(\psi^k_\alpha,\phi^l)\Big]\\[3pt]
&\kern.5cm-2R^{ijkl}_{(2)}\Big[(\bar\psi_{i\alpha },\bar\psi^\alpha_j)(\bphi_k,\bphi_l)+
  2(\bar\psi_{i\alpha},\bphi_j)(\bar\psi_k^\alpha,\bphi_l)\Big]\\[3pt]
&\kern.5cm-16\left[R_{ijkl}^{(1)}R^{ij'k'l'}_{(2)}([\phi^l,\phi^k,\phi^j],\bar\phi_{j'})
     (\bar\phi_{l'},\bar\phi_{k'})\right.\\[3pt]
&\kern2cm + R_{ijkl}^{(2)}R^{ij'k'l'}_{(1)}([\bar\phi_{l'},\bar\phi_{k'},\bar\phi_{j'}],\phi^j)
     (\phi^l,\phi^k)\\[3pt]
&\kern2cm\left.+\,R_{ijkl}^{(2)}R^{ij'k'l'}_{(2)}(\phi^l,\phi^k)(\bar\phi_{l'},\bar\phi_{k'})
       (\phi^j,\bar\phi_{j'})\right].
\end{aligned}
\end{equation}

\bigskip
\noindent{\it \underline{Component action in the Hermitian case:}}
\bigskip

\noindent 
Let us now discuss the Hermitian case. Here, we shall assume that $\beta=0$, i.e.\ we work with
the usual Hermitian 3-bracket in the superpotential. In terms of the component fields
\eqref{eq:WZGaugeComponents}, the action \eqref{eq:S0Complex} reads as
\begin{equation}\label{eq:S0H}
\begin{aligned}
S^H_0\ &=\ \int \dd^3 x\, \bigg[
  \eps^{\mu\nu\lambda}\lbr A_\mu,\dpar_\nu A_\lambda+\tfrac{1}{3\sqrt{\kappa}}\lsb A_\nu,A_\lambda\rsb\rbr
  -\di \lbr\lambda_\alpha,\bar\lambda^\alpha\rbr-\di\lbr\bl_\alpha,\lambda^\alpha\rbr
  -\lbr D,\sigma\rbr-\lbr \sigma,D\rbr\\
 &\kern1cm
  +(F^m,F^m)+(\bar{F}_\mdt,\bar{F}_\mdt)
  -(\nabla_\mu \phi^m,\nabla^\mu \phi^m)
  -(\nabla_\mu \bar{\phi}_\mdt,\nabla^\mu \bar{\phi}_\mdt)
  -\di(\psi^{m\alpha},\nabla_{\alpha\beta}\psi^{m\beta})\\
 &\kern1.5cm
  +\di(\bar{\psi}^{m\alpha},\nabla_{\alpha\beta}\bar{\psi}^{m\beta})
  -\tfrac{\di}{\sqrt{\kappa}}(\phi^m,D(\phi^m))
  +\tfrac{\di}{\sqrt{\kappa}}(\bar{\phi}_\mdt,D(\bar{\phi}_\mdt))
  +\sqrt{\tfrac{2}{\kappa}}(\phi^m,\lambda^\alpha(\psi^m_\alpha))\\
 &\kern2cm
  -\sqrt{\tfrac{2}{\kappa}}(\bpsi_\mdt^{\alpha},\lambda_\alpha(\bphi_\mdt))
  +\sqrt{\tfrac{2}{\kappa}}(\lambda^\alpha(\psi^m_\alpha),\phi^m)\\
 &\kern2.5cm
  -\sqrt{\tfrac{2}{\kappa}}(\lambda^\alpha(\bphi_\mdt),\bpsi_{\mdt\alpha})
  +\tfrac{1}{\kappa}(\phi^m,\sigma^2(\phi^m))\\
 &\kern3cm
  +\tfrac{1}{\kappa}(\bphi_\mdt,\sigma^2(\bphi_\mdt))
  +\tfrac{1}{\sqrt{\kappa}}(\psi^m_\alpha,\sigma(\psi^{m\alpha}))
  -\tfrac{1}{\sqrt{\kappa}}(\bpsi_{\mdt}^\alpha,\sigma(\bpsi_\mdt^{\alpha}))\bigg]~.
\end{aligned}
\end{equation}
In component form, the superpotential terms \eqref{eq:S1Complex} are given by
\begin{equation}\label{eq:S1H}
\begin{aligned}
   S_1^H\ &=\ -2\int \dd^3x~\bigg\{ 
   H_{mn\mdt\ndt}^{(1)}\bigg[
    (\bar F_\mdt,[\phi^m,\phi^n;\bar\phi_\ndt])+([\bar\phi_\mdt,\bar\phi_\ndt;\phi^n],F^m)\\
   &\kern3cm
    +(\bar\psi_{\ndt\alpha},[\psi^{m\alpha},\phi^n;\bar\phi_\mdt])
    +(\bar\phi_\ndt,[\psi^m_\alpha,\phi^n;\bar\psi_\mdt^\alpha])\\
   &\kern3.5cm
    -\tfrac12(\bar\psi_{\ndt\alpha},[\phi^m,\phi^n;\bar\psi^\alpha_\mdt])
    -\tfrac12(\bar\phi_\ndt,[\psi^m_\alpha,\psi^{n\alpha};\bar\phi_\mdt)\bigg]\\
   &\kern2.2cm+H_{mn\mdt\ndt}^{(2)}\bigg[
    -(\bar F_\mdt,\phi^m)(\bar\phi_\ndt,\phi^n)-(\bar\phi_\mdt,F^m)(\bar\phi_\ndt,\phi^n)\\
   &\kern3cm
    +(\bar\psi_{\mdt\alpha},\psi^{m\alpha})(\bar\phi_\ndt,\phi^n)
    +(\bar\psi_{\mdt\alpha},\phi^m)(\bar\phi_\ndt,\psi^{n\alpha})\\
   &\kern3.5cm
    -\tfrac12(\bar\psi_{\mdt\alpha},\phi^m)(\bar\psi_{\ndt}^\alpha,\phi^n)
    -\tfrac12(\bar\phi_\mdt,\psi^m_\alpha)(\bar\phi_\ndt,\psi^{n\alpha})\bigg]\bigg\}
+\mathrm{c.c.}~.
\end{aligned} 
\end{equation}
Varying $S^H=S_0^H+S_1^H$, we find the following (algebraic) equations
of motion for the auxiliary fields $F^m$, $\bar F_m$, $F^\mdt$, $\bar F_\mdt$, $D$, $\sigma$, $\lambda$ and $\bar\lambda$:
\begin{equation}\label{eq:AuxEqComplex}
 \begin{aligned}
   F^m\ &=\ 2 H^{\mdt\ndt mn}_{(1)}[\bphi_\mdt,\bphi_\ndt;\phi^n]
            -2 H^{\mdt\ndt mn}_{(2)}\bphi_\mdt(\phi^n,\bphi_\ndt)~,\\
  \bar F_m\ &=\ -2H_{mn\mdt\ndt}^{(1)}[\phi^\mdt,\phi^\ndt;\bphi_n]
                   -2H_{mn\mdt\ndt}^{(2)}\phi^\mdt(\bphi_\ndt,\phi^n)~,\\
  F^\mdt\ &=\ -2 H^{\mdt\ndt mn}_{(1)}[\bphi_m,\bphi_n;\phi^\mdt]
            -2 H^{\mdt\ndt mn}_{(2)}\bphi_m(\phi^\ndt,\bphi_n)~,\\
  \bar F_\mdt\ &=\ 2H_{mn\mdt\ndt}^{(1)}[\phi^m,\phi^n;\bphi_\ndt]
                   -2H_{mn\mdt\ndt}^{(2)}\phi^m(\bphi_n,\phi^\ndt)~,\\
  D(A)\ &=\ \tfrac{1}{2\kappa}\Big([A,\sigma(\phi^m);\phi^m]
             +[A,\sigma(\bphi_\mdt);\bphi_\mdt]
             -[A,\phi^m;\sigma(\phi^m)]
             -[A,\bphi_\mdt;\sigma(\bphi_\mdt)]\Big)\\
       &\kern3cm
         -\tfrac{1}{2\sqrt{\kappa}}\Big([A,\psi^{m\alpha};\psi^{m}_\alpha]
         -[A,\bpsi_\mdt^{\alpha};\bpsi_{\mdt\alpha}]\Big),\\
  \sigma(A)\ &=\ 
    -\tfrac{\di}{2\sqrt{\kappa}}\left([A,\phi^m;\phi^m]-[A,\bphi_{\mdt};\bphi_\mdt]\right),\\
  \lambda_\alpha(A)\ &=\ \tfrac{\di(-1)^{\tilde{A}}}{\sqrt{2\kappa}}
      \left([A,\phi^m;\psi^{m}_\alpha]-[A,\bpsi_{\mdt\alpha};\bphi_{\mdt}]\right),\\
  \bar\lambda_\alpha(A)\ &=\ \tfrac{\di(-1)^{\tilde{A}}}{\sqrt{2\kappa}}
     \left([A,\psi^m_\alpha;\phi^m]-[A,\bphi_\mdt;\bpsi_{\mdt\alpha}]\right),
 \end{aligned}
\end{equation}
where $A$ is an arbitrary field taking values in $\CA$. We may now substitute these
expressions into equations \eqref{eq:S0H} and \eqref{eq:S1H} to arrive at the final expression
for the component action. Since this is a rather lengthy expression and moreover basically of the same
form as \eqref{eq:S01RComp}, we shall not display the full action here but only give the multi-trace terms:
\begin{equation}
 \begin{aligned}
   \CL^H_{mult}\ &=\ 
     -2 H_{mn\mdt\ndt}^{(2)}\Big[
    (\bar\psi_{\mdt\alpha},\psi^{m\alpha})(\bar\phi_\ndt,\phi^n)
    +(\bar\psi_{\mdt\alpha},\phi^m)(\bar\phi_\ndt,\psi^{n\alpha})\\
   &\kern3cm
    -\tfrac12(\bar\psi_{\mdt\alpha},\phi^m)(\bar\psi_{\ndt}^\alpha,\phi^n)
    -\tfrac12(\bar\phi_\mdt,\psi^m_\alpha)(\bar\phi_\ndt,\psi^{n\alpha})\Big]\\
     &\kern.5cm
  -2 H^{\mdt\ndt mn}_{(2)}\Big[
    (\psi^{m\alpha},\bar\psi_{\mdt\alpha})(\phi^n,\bar\phi_\ndt)
    +(\psi^{n\alpha},\bar\phi_\ndt)(\phi^m,\bar\psi_{\mdt\alpha})\\
   &\kern3cm
    -\tfrac12(\phi^n,\bar\psi_{\ndt}^\alpha)(\phi^m,\bar\psi_{\mdt\alpha})
    -\tfrac12(\psi^{n\alpha},\bar\phi_\ndt)(\psi^m_\alpha,\bar\phi_\mdt)\Big]\\
   &\kern.5cm
    +4H_{mn\mdt\ndt}^{(1)}H^{\mdt'\ndt'mn'}_{(2)}
       ([\bphi_\mdt,\bphi_\ndt;\phi^n],\bphi_{\mdt'})(\phi^{n'},\bphi_{\ndt'})\\
   &\kern.5cm
    +4H_{mn\mdt\ndt}^{(2)}H^{\mdt'\ndt'mn'}_{(1)}
       (\bphi_\mdt,[\bphi_{\mdt'},\bphi_{\ndt'};\phi^{n'}])(\bphi_\ndt,\phi^n)\\
   &\kern.5cm
    -4H_{mn\mdt\ndt}^{(2)}H^{\mdt'\ndt'mn'}_{(2)}
       (\bphi_\mdt,\bphi_{\mdt'})(\bphi_\ndt,\phi^n)(\phi^{n'},\bphi_{\ndt'})\\
   &\kern.5cm
    +4H^{\mdt\ndt mn}_{(1)}H_{m'n'\mdt\ndt'}^{(2)}
      ([\phi^m,\phi^n;\bphi_\ndt],\phi^{m'})(\bphi_{\ndt'},\phi^{n'})\\
   &\kern.5cm
    +4H^{\mdt\ndt mn}_{(2)}H_{m'n'\mdt\ndt'}^{(1)}
       (\phi^m,[\phi^{m'},\phi^{n'};\bphi_{\ndt'}])(\phi^n,\bphi_\ndt)\\
   &\kern.5cm
    -4H^{\mdt\ndt mn}_{(2)}H_{m'n'\mdt\ndt'}^{(2)}
       (\phi^m,\phi^{m'})(\phi^n,\bphi_\ndt)(\bphi_{\ndt'},\phi^{n'})~.
 \end{aligned}
\end{equation}

\subsection{Feynman rules: Vertices}\label{app:vertices}

\bigskip
\noindent{\it \underline{Vertices for real 3-algebras:}}
\bigskip

\noindent 
Let us list the Feynman rules for the vertices in Landau gauge $\alpha\beta\to0$. They are:

\begin{subequations}\label{eq:Vertex}
\vspace*{-40pt}
\begin{eqnarray}
\kern-1cm\mbox{$V^3$-vertex :}&&\kern30pt
  \begin{picture}(70,70)(0,20)
   \SetScale{1}
   \put(0,0){
   \Vertex(25,25){1}
    \Photon(25,25)(60.36,25){3}{6}
    \Photon(0,0)(25,25){-3}{6}
    \Photon(25,25)(0,50){-3}{6}
    \Text(63.36,25)[l]{$A_1$}
    \Text(0,52)[br]{$A_2$}
    \Text(0,-2)[tr]{$A_3$}
    \Text(1,50)[bl]{$\searrow k_2$}
    \Text(4,4)[tl]{$\nearrow k_3$}
    \Text(40,33)[tl]{$\longleftarrow$}
    \Text(25,45)[tl]{$-k_2-k_3$}
    \Text(25,15)[l]{$\theta$}
   }
  \end{picture}
  \kern20pt=\kern20pt \tfrac{2\di}{\sqrt{\kappa}}V_{A_1A_2A_3}~,\label{eq:VVV-Vertex} \\
\kern-1cm\mbox{$\Phi V \Phib$-vertex :}&&\kern30pt
  \begin{picture}(70,70)(0,20)
   \SetScale{1}
   \put(0,0){
   \Vertex(25,25){1}
    \Photon(25,25)(60.36,25){3}{6}
    \ArrowLine(0,0)(25,25)
    \ArrowLine(25,25)(0,50)
    \Text(63.36,25)[l]{$A$}
     \Text(0,52)[br]{$J$}
    \Text(0,-2)[tr]{$I$}
\Text(25,15)[l]{$\theta$}
   }
  \end{picture}
  \kern20pt=\kern20pt \di \tfrac{-2\di}{\sqrt{\kappa}} {T_{AI}}^J~,\label{eq:VPP-Vertex}\\
\kern-1cm\mbox{$\Phi V^2 \Phib$-vertex :}&&\kern30pt
  \begin{picture}(70,70)(0,20)
   \SetScale{1}
    \put(0,0){
    \Vertex(25,25){1}
    \Photon(25,25)(50,0){3}{6}
    \Photon(25,25)(50,50){-3}{6}
    \ArrowLine(0,0)(25,25)
    \ArrowLine(25,25)(0,50)
     \Text(0,52)[br]{$J$}
    \Text(0,-2)[tr]{$I$}
    \Text(52,-2)[tl]{$A$}
    \Text(52,52)[bl]{$B$}
\Text(22,15)[l]{$\theta$}
   }
  \end{picture}
  \kern20pt=\kern20pt {\di}\big(\tfrac{-2\di}{\sqrt{\kappa}}\big)^2 {T_{(AI}}^K{T_{B)K}}^J~,
  \label{eq:VPPPP-Vertex}\\
\kern-1cm\mbox{$\Phi^4$-vertex :}&&\kern30pt
  \begin{picture}(70,70)(0,20)
   \SetScale{1}
   \put(0,0){
    \Vertex(25,25){1}
    \ArrowLine(50,0)(25,25)
    \ArrowLine(50,50)(25,25)
    \ArrowLine(0,0)(25,25)
    \ArrowLine(0,50)(25,25)
\Text(25,15)[l]{$\theta$}
 \Text(0,52)[br]{$L$}
    \Text(0,-2)[tr]{$I$}
    \Text(52,-2)[tl]{$J$}
    \Text(52,52)[bl]{$K$}
   }
  \end{picture}
  \kern20pt=\kern20pt \di 4! R_{IJKL}~,\label{eq:PPPP-Vertex}\\[20pt]\notag
\end{eqnarray}

\vspace*{-75pt}
\begin{eqnarray}
\kern-1cm\mbox{$\Phib^4$-vertex :}&&\kern30pt
  \begin{picture}(70,70)(0,20)
   \SetScale{1}
   \put(0,0){
    \Vertex(25,25){1}
    \ArrowLine(25,25)(50,0)
    \ArrowLine(25,25)(50,50)
    \ArrowLine(25,25)(0,0)
    \ArrowLine(25,25)(0,50)
\Text(0,52)[br]{$L$}
    \Text(0,-2)[tr]{$I$}
    \Text(52,-2)[tl]{$J$}
    \Text(52,52)[bl]{$K$}
\Text(25,15)[l]{$\theta$}
   }
  \end{picture}
  \kern20pt=\kern20pt \di 4! R^{IJKL}~,\label{eq:PPPPbar-Vertex}\\
\kern-1cm\mbox{ghost/gluon-vertices :}&&\kern30pt
  \begin{picture}(70,70)(0,20)
   \SetScale{1}
    \put(0,0){
    \Vertex(25,25){1}
     \Photon(25,25)(60.36,25){3}{6}
     \DashLine(0,0)(25,25){4}
     \DashLine(25,25)(0,50){4}
     \Text(63.36,25)[l]{$A$}
     \Text(0,52)[br]{$B$}
     \Text(0,-2)[tr]{$C$}
\Text(25,15)[l]{$\theta$}
    }
   \end{picture}
   \kern20pt=\kern20pt \di(-1)^{\#}\tfrac{\di}{\sqrt{\kappa}}F_{ABC}~.\label{eq:VG-Vertex}
\end{eqnarray}
\vspace*{10pt}
\end{subequations}

\noindent
Here, `$\#$' is the number of antichiral (anti)ghosts entering the vertex. We have not put arrows on the ghost lines, since we have vertices where only either chiral or antichiral ghosts enter or where chiral and antichiral ghosts enter. Furthermore,
\begin{subequations}
\begin{equation}
 V_{A_1A_2A_3}\ =\ \sum_{r,s}F_{A_1A_rA_s}[\Db(-k_r,\theta)\Delta_{r\theta}(k_r)][D(-k_s,\theta)\Delta_{s\theta}(k_s)]
\end{equation}
with
\begin{equation}
 \Delta_{ij}(k_i)\ :=\ -\tfrac{\di}{4k_i^2}\Db D(k_i,\theta_i)\delta_{ij}\eand
 \delta_{ij}\ :=\delta^{(4)}(\theta_i-\theta_j)~.
\end{equation}
\end{subequations}
The coefficients appearing in \eqref{eq:PPPP-Vertex} and \eqref{eq:PPPPbar-Vertex} are
\begin{equation}\label{eq:DefOfSymR}
\begin{aligned}
 R_{IJKL}\ &=\ \big[R_{ijkl}^{(1)}f_{abcd}+R_{ijkl}^{(2)}h_{ab}h_{cd}\big]_s\\          
           &=\ \tfrac{1}{3}\big(R_{ijkl}^{(1)}f_{abcd}+R_{iklj}^{(1)}f_{acdb}
            +R_{iljk}^{(1)}f_{adbc}+R_{ijkl}^{(2)}h_{ab}h_{cd}+R_{iklj}^{(2)}h_{ac}h_{db}
            +R_{iljk}^{(2)}h_{ad}h_{bc}\big)~,\\
 R^{IJKL}\ &=\ \big[R^{ijkl}_{(1)}f^{abcd}+R^{ijkl}_{(2)}h^{ab}h^{cd}\big]_s\\          
           &=\ \tfrac{1}{3}\big(R^{ijkl}_{(1)}f^{abcd}+R^{iklj}_{(1)}f^{acdb}
            +R^{iljk}_{(1)}f^{adbc}+R^{ijkl}_{(2)}h^{ab}h^{cd}+R^{iklj}_{(2)}h^{ac}h^{db}
            +R^{iljk}_{(2)}h^{ad}h^{bc}\big)~.
\end{aligned}
\end{equation}
The subscript `$s$' refers to total symmetrization in the multi-indices $IJKL$.

\bigskip
\noindent{\it \underline{Vertices for Hermitian 3-algebras:}}
\bigskip

\noindent 
In the Hermitian case, the Feynman rules for the vertices are very similar
to the ones for real 3-algebras. The purely gluonic and gluon/ghost vertices
are the same and we shall again adopt Landau gauge.
 The only difference is in the gluon/matter and pure matter vertices,
since we have two different types of matter: $\Phi^I$ and $\Phi_{\dot I}$. We have

\begin{subequations}\label{eq:VertexComplex}
\vspace*{-40pt}
\begin{eqnarray}
\kern-1cm\mbox{$\Phi V \Phib$-vertex :}&&\kern30pt
  \begin{picture}(70,70)(0,20)
   \SetScale{1}
   \put(0,0){
   \Vertex(25,25){1}
    \Photon(25,25)(60.36,25){3}{6}
    \ArrowLine(0,0)(25,25)
    \ArrowLine(25,25)(0,50)
    \Text(63.36,25)[l]{$A$}
     \Text(0,52)[br]{$J$}
    \Text(0,-2)[tr]{$I$}
\Text(25,15)[l]{$\theta$}
   }
  \end{picture}
  \kern20pt=\kern20pt \di \tfrac{-2\di}{\sqrt{\kappa}} {T_{AI}}^J~,\\
\kern-1cm\mbox{$\Phi V \Phib$-vertex :}&&\kern30pt
  \begin{picture}(70,70)(0,20)
   \SetScale{1}
   \put(0,0){
   \Vertex(25,25){1}
    \Photon(25,25)(60.36,25){3}{6}
    \ArrowLine(0,0)(25,25)
    \ArrowLine(25,25)(0,50)
    \Text(63.36,25)[l]{$A$}
     \Text(0,52)[br]{$\dot J$}
    \Text(0,-2)[tr]{$\dot I$}
\Text(25,15)[l]{$\theta$}
   }
  \end{picture}
  \kern20pt=\kern20pt \di \tfrac{2\di}{\sqrt{\kappa}} {T_{A\dot I}}^{\dot J}~,\\[20pt]\notag
\end{eqnarray}
\vspace*{20pt}

\vspace*{-90pt}
\begin{eqnarray}
\kern-1cm\mbox{$\Phi V^2 \Phib$-vertex :}&&\kern30pt
  \begin{picture}(70,70)(0,20)
   \SetScale{1}
    \put(0,0){
    \Vertex(25,25){1}
    \Photon(25,25)(50,0){3}{6}
    \Photon(25,25)(50,50){-3}{6}
    \ArrowLine(0,0)(25,25)
    \ArrowLine(25,25)(0,50)
     \Text(0,52)[br]{$J$}
    \Text(0,-2)[tr]{$I$}
    \Text(52,-2)[tl]{$A$}
    \Text(52,52)[bl]{$B$}
\Text(22,15)[l]{$\theta$}
   }
  \end{picture}
  \kern20pt=\kern20pt {\di}\big(\tfrac{-2\di}{\sqrt{\kappa}}\big)^2 {T_{(AI}}^K{T_{B)K}}^J~,\\
\kern-1cm\mbox{$\Phi V^2 \Phib$-vertex :}&&\kern30pt
  \begin{picture}(70,70)(0,20)
   \SetScale{1}
    \put(0,0){
    \Vertex(25,25){1}
    \Photon(25,25)(50,0){3}{6}
    \Photon(25,25)(50,50){-3}{6}
    \ArrowLine(0,0)(25,25)
    \ArrowLine(25,25)(0,50)
     \Text(0,52)[br]{$\dot J$}
    \Text(0,-2)[tr]{$\dot I$}
    \Text(52,-2)[tl]{$A$}
    \Text(52,52)[bl]{$B$}
\Text(22,15)[l]{$\theta$}
   }
  \end{picture}
  \kern20pt=\kern20pt {\di}\big(\tfrac{2\di}{\sqrt{\kappa}}\big)^2 
{T_{(A\dot I}}^{\dot K}{T_{B)\dot K}}^{\dot J}~,\\
\kern-1cm\mbox{$\Phi^4$-vertex :}&&\kern30pt
  \begin{picture}(70,70)(0,20)
   \SetScale{1}
   \put(0,0){
    \Vertex(25,25){1}
    \ArrowLine(50,0)(25,25)
    \ArrowLine(50,50)(25,25)
    \ArrowLine(0,0)(25,25)
    \ArrowLine(0,50)(25,25)
\Text(25,15)[l]{$\theta$}
 \Text(0,52)[br]{$\dot L$}
    \Text(0,-2)[tr]{$I$}
    \Text(52,-2)[tl]{$J$}
    \Text(52,52)[bl]{$\dot K$}
   }
  \end{picture}
  \kern20pt=\kern20pt \di 4 {H_{IJ}}^{\dot K\dot L}~,\\
\kern-1cm\mbox{$\Phib^4$-vertex :}&&\kern30pt
  \begin{picture}(70,70)(0,20)
   \SetScale{1}
   \put(0,0){
    \Vertex(25,25){1}
    \ArrowLine(25,25)(50,0)
    \ArrowLine(25,25)(50,50)
    \ArrowLine(25,25)(0,0)
    \ArrowLine(25,25)(0,50)
\Text(0,52)[br]{$\dot L$}
    \Text(0,-2)[tr]{$I$}
    \Text(52,-2)[tl]{$J$}
    \Text(52,52)[bl]{$\dot K$}
\Text(25,15)[l]{$\theta$}
   }
  \end{picture}
  \kern20pt=\kern20pt \di {4 H_{\dot K\dot L}}^{IJ}~,
\end{eqnarray}
\end{subequations}

\vspace*{30pt}
\noindent
where
\begin{equation}
 \begin{aligned}
  {H_{IJ}}^{\dot K\dot L}\ &=\ \big[H_{mn\mdt\ndt}^{(1)}{g_{ab}}^{cd}+H_{mn\mdt\ndt}^{(2)}
       \delta_a^{~c}\delta_b^{~d}\big]_s\\ 
  &=\ \tfrac12\big[H_{mn\mdt\ndt}^{(1)}{g_{ab}}^{cd}+H_{mn\ndt\mdt}^{(1)}{g_{ab}}^{dc}+H_{mn\mdt\ndt}^{(2)}
       \delta_a^{~c}\delta_b^{~d}+H_{mn\ndt\mdt}^{(2)}
       \delta_a^{~d}\delta_b^{~c}\big]~,\\
  {H_{\dot K\dot L}}^{IJ}\ &=\ \big[H^{\ndt\mdt mn}_{(1)}{g_{dc}}^{ab}+H^{\mdt\ndt mn}_{(2)}
         \delta_c^{~a}\delta_d^{~b}\big]_s\\ 
  &=\ \tfrac12\big[H^{\ndt\mdt mn}_{(1)}{g_{dc}}^{ab}+H^{\mdt\ndt mn}_{(1)}{g_{cd}}^{ab}+
H^{\mdt\ndt mn}_{(2)}
         \delta_c^{~a}\delta_d^{~b}+H^{\ndt\mdt mn}_{(2)}
         \delta_d^{~a}\delta_c^{~b}\big]~,\\
 \end{aligned}
\end{equation}
where `$s$' refers again to total symmetrization.


\noindent


\bigskip
\bigskip

\begin{thebibliography}{10}
\ifx\href\asklfhas\newcommand{\href}[2]{#2}\fi
\ifx\arxivref\asklfhas\newcommand{\arxivref}[2]{\href{http://arxiv.org/abs/#1}%
{#2}}\fi
\ifx\doiref\asklfhas\newcommand{\doiref}[2]{\href{http://dx.doi.org/#1}{#2}}\fi
\raggedright
\small
\parskip 0pt

%%CITATION = HEP-TH/0611108;%%
\bibitem{Bagger:2006sk}
J.~Bagger and N.~Lambert,
\textit{``Modeling multiple M2's''},
\textsf{\doiref{10.1103/PhysRevD.75.045020}{Phys.~Rev.~D75,~045020~(2007)}},
\texttt{\arxivref{hep-th/0611108}{hep-th/0611108}}.

%%CITATION = 0711.0955;%%
\bibitem{Bagger:2007jr}
J.~Bagger and N.~Lambert,
\textit{``Gauge symmetry and supersymmetry of multiple M2-branes''},
\textsf{\doiref{10.1103/PhysRevD.77.065008}{Phys.~Rev.~D77,~065008~(2008)}},
\texttt{\arxivref{0711.0955}{arxiv:0711.0955}}.

%%CITATION = 0709.1260;%%
\bibitem{Gustavsson:2007vu}
A.~Gustavsson,
\textit{``Algebraic structures on parallel M2-branes''},
\textsf{\doiref{10.1016/j.nuclphysb.2008.11.014}{Nucl.~Phys.~B811,~66~(2009)}},
\texttt{\arxivref{0709.1260}{arxiv:0709.1260}}.

%%CITATION = 0901.3905;%%
\bibitem{Lazaroiu:2009wz}
C.~I.~Lazaroiu, D.~McNamee, C.~Saemann and A.~Zejak,
\textit{``{Strong homotopy Lie algebras, generalized Nahm equations and
  multiple M2-branes}''},
\texttt{\arxivref{0901.3905}{arxiv:0901.3905}}.

\bibitem{Filippov:1985aa}
V.~T.~Filippov,
\textit{``$n$-Lie algebras''},
\textsf{Sib.~Mat.~Zh.~26,~126~(1985)}.

\bibitem{Nagy:2007aa}
P.-A.~Nagy,
\textit{``Prolongations of Lie algebras and applications''},
\texttt{\arxivref{0712.1398}{arxiv:0712.1398}}.

%%CITATION = 0804.2662;%%
\bibitem{Papadopoulos:2008sk}
G.~Papadopoulos,
\textit{``{M2-branes, 3-Lie algebras and Pl{\"u}cker relations}''},
\textsf{\doiref{10.1088/1126-6708/2008/05/054}{JHEP~0805,~054~(2008)}},
\texttt{\arxivref{0804.2662}{arxiv:0804.2662}}.

 
%%CITATION = 0804.3078;%%
\bibitem{Gauntlett:2008uf}
J.~P.~Gauntlett and J.~B.~Gutowski,
\textit{``{Constraining maximally supersymmetric membrane actions}''},
\textsf{\doiref{10.1088/1126-6708/2008/06/053}{JHEP~0806,~053~(2008)}},
\texttt{\arxivref{0804.3078}{arxiv:0804.3078}}.

%%CITATION = 0807.0163;%%
\bibitem{Bagger:2008se}
J.~Bagger and N.~Lambert,
\textit{``{Three-algebras and $\CN=6$ Chern-Simons gauge theories}''},
\textsf{\doiref{10.1103/PhysRevD.79.025002}{Phys.~Rev.~D79,~025002~(2009)}},
\texttt{\arxivref{0807.0163}{arxiv:0807.0163}}.

%%CITATION = 0807.0808;%%
\bibitem{Cherkis:2008qr}
S.~Cherkis and C.~Saemann,
\textit{``{Multiple M2-branes and generalized 3-Lie algebras}''},
\textsf{\doiref{10.1103/PhysRevD.78.066019}{Phys.~Rev.~D78,~066019~(2008)}},
\texttt{\arxivref{0807.0808}{arxiv:0807.0808}}.

%%CITATION = 0806.1218;%%
\bibitem{Aharony:2008ug}
O.~Aharony, O.~Bergman, D.~L.~Jafferis and J.~Maldacena,
\textit{``{$\CN=6$ superconformal Chern-Simons-matter theories, M2-branes and
  their gravity duals}''},
\textsf{\doiref{10.1088/1126-6708/2008/10/091}{JHEP~0810,~091~(2008)}},
\texttt{\arxivref{0806.1218}{arxiv:0806.1218}}.

%%CITATION = 0803.3803;%%
\bibitem{VanRaamsdonk:2008ft}
M.~Van~Raamsdonk,
\textit{``{Comments on the Bagger-Lambert theory and multiple M2-branes}''},
\textsf{\doiref{10.1088/1126-6708/2008/05/105}{JHEP~0805,~105~(2008)}},
\texttt{\arxivref{0803.3803}{arxiv:0803.3803}}.

%%CITATION = 0806.3951;%%
\bibitem{Minahan:2008hf}
J.~A.~Minahan and K.~Zarembo,
\textit{``{The Bethe ansatz for superconformal Chern-Simons}''},
\textsf{\doiref{10.1088/1126-6708/2008/09/040}{JHEP~0809,~040~(2008)}},
\texttt{\arxivref{0806.3951}{arxiv:0806.3951}}.

%%CITATION = 0806.4589;%%
\bibitem{Gaiotto:2008cg}
D.~Gaiotto, S.~Giombi and X.~Yin,
\textit{``{Spin chains in $\CN=6$ superconformal Chern-Simons-matter
  theory}''},
\textsf{\doiref{10.1088/1126-6708/2009/04/066}{JHEP~0904,~066~(2009)}},
\texttt{\arxivref{0806.4589}{arxiv:0806.4589}}.

%%CITATION = 0807.0777;%%
\bibitem{Gromov:2008qe}
N.~Gromov and P.~Vieira,
\textit{``{The all loop AdS4/CFT3 Bethe ansatz}''},
\textsf{\doiref{10.1088/1126-6708/2009/01/016}{JHEP~0901,~016~(2009)}},
\texttt{\arxivref{0807.0777}{arxiv:0807.0777}}.

%%CITATION = 0807.2063;%%
\bibitem{Bak:2008cp}
D.~Bak and S.-J.~Rey,
\textit{``{Integrable spin chain in superconformal Chern-Simons theory}''},
\textsf{\doiref{10.1088/1126-6708/2008/10/053}{JHEP~0810,~053~(2008)}},
\texttt{\arxivref{0807.2063}{arxiv:0807.2063}}.

%%CITATION = 0901.0411;%%
\bibitem{Zwiebel:2009vb}
B.~I.~Zwiebel,
\textit{``{Two-Loop integrability of planar $\CN=6$ superconformal Chern-Simons
  theory}''},
\textsf{\doiref{10.1088/1751-8113/42/49/495402}{J.~Phys.~A42,~495402~(2009)}},
\texttt{\arxivref{0901.0411}{arxiv:0901.0411}}.

%%CITATION = 0901.1142;%%
\bibitem{Minahan:2009te}
J.~A.~Minahan, W.~Schulgin and K.~Zarembo,
\textit{``{Two loop integrability for Chern-Simons theories with $\CN=6$
  supersymmetry}''},
\textsf{\doiref{10.1088/1126-6708/2009/03/057}{JHEP~0903,~057~(2009)}},
\texttt{\arxivref{0901.1142}{arxiv:0901.1142}}.

%%CITATION = 0904.4677;%%
\bibitem{Bak:2009mq}
D.~Bak, H.~Min and S.-J.~Rey,
\textit{``{Generalized dynamical spin chain and 4-loop integrability in $\CN=6$
  superconformal Chern-Simons theory}''},
\textsf{\doiref{10.1016/j.nuclphysb.2009.10.011}{Nucl.~Phys.~B827,~381~(2010)}},
\texttt{\arxivref{0904.4677}{arxiv:0904.4677}}.

%%CITATION = 0704.3740;%%
\bibitem{Gaiotto:2007qi}
D.~Gaiotto and X.~Yin,
\textit{``{Notes on superconformal Chern-Simons-matter theories}''},
\textsf{\doiref{10.1088/1126-6708/2007/08/056}{JHEP~0708,~056~(2007)}},
\texttt{\arxivref{0704.3740}{arxiv:0704.3740}}.

%%CITATION = HEP-TH 9503121;%%
\bibitem{Leigh:1995ep}
R.~G.~Leigh and M.~J.~Strassler,
\textit{``Exactly marginal operators and duality in four-dimensional $\CN=1$
  supersymmetric gauge theory''},
\textsf{\doiref{10.1016/0550-3213(95)00261-P}{Nucl.~Phys.~B447,~95~(1995)}},
\texttt{\arxivref{hep-th/9503121}{hep-th/9503121}}.

%%CITATION = 0803.3218;%%
\bibitem{Mukhi:2008ux}
S.~Mukhi and C.~Papageorgakis,
\textit{``{M2 to D2}''},
\textsf{\doiref{10.1088/1126-6708/2008/05/085}{JHEP~0805,~085~(2008)}},
\texttt{\arxivref{0803.3218}{arxiv:0803.3218}}.

%%CITATION = 0803.3611;%%
\bibitem{Berman:2008be}
D.~S.~Berman, L.~C.~Tadrowski and D.~C.~Thompson,
\textit{``Aspects of multiple membranes''},
\textsf{\doiref{10.1016/j.nuclphysb.2008.05.006}{Nucl.~Phys.~B802,~106~(2008)}%
},
\texttt{\arxivref{0803.3611}{arxiv:0803.3611}}.

%%CITATION = 0808.1271;%%
\bibitem{Imeroni:2008cr}
E.~Imeroni,
\textit{``{On deformed gauge theories and their string/M-theory duals}''},
\textsf{\doiref{10.1088/1126-6708/2008/10/026}{JHEP~0810,~026~(2008)}},
\texttt{\arxivref{0808.1271}{arxiv:0808.1271}}.

%%CITATION = 0905.0709;%%
\bibitem{Craps:2009qc}
B.~Craps, T.~Hertog and N.~Turok,
\textit{``{A multitrace deformation of ABJM theory}''},
\textsf{\doiref{10.1103/PhysRevD.80.086007}{Phys.~Rev.~D80,~086007~(2009)}},
\texttt{\arxivref{0905.0709}{arxiv:0905.0709}}.

%%CITATION = 0809.1086;%%
\bibitem{deMedeiros:2008zh}
P.~de~Medeiros, J.~Figueroa-O'Farrill, E.~Mendez-Escobar and P.~Ritter,
\textit{``{On the Lie-algebraic origin of metric 3-algebras}''},
\textsf{\doiref{10.1007/s00220-009-0760-1}{Commun.~Math.~Phys.~290,~871~(2009)%
}},
\texttt{\arxivref{0809.1086}{arxiv:0809.1086}}.

%%CITATION = 0812.3127;%%
\bibitem{Cherkis:2008ha}
S.~Cherkis, V.~Dotsenko and C.~Saemann,
\textit{``{On superspace actions for multiple M2-branes, metric 3-algebras and
  their classification}''},
\textsf{\doiref{10.1103/PhysRevD.79.086002}{Phys.~Rev.~D79,~086002~(2009)}},
\texttt{\arxivref{0812.3127}{arxiv:0812.3127}}.

\bibitem{Wess:1992cp}
J.~Wess and J.~Bagger,
\textit{``Supersymmetry and supergravity''},
Princeton, USA: Univ. Pr. (1992).

%%CITATION = TMPHA,77,1070;%%
\bibitem{Zupnik:1988en}
B.~M.~Zupnik and D.~G.~Pak,
\textit{``{Superfield formulation of the simplest three-dimensional gauge
  theories and conformal supergravities}''},
\textsf{\doiref{10.1007/BF01028682}{Theor.~Math.~Phys.~77,~1070~(1988)}}.

%%CITATION = PHLTA,B268,203;%%
\bibitem{Ivanov:1991fn}
E.~A.~Ivanov,
\textit{``{Chern-Simons matter systems with manifest $\CN=2$ supersymmetry}''},
\textsf{\doiref{10.1016/0370-2693(91)90804-Y}{Phys.~Lett.~B268,~203~(1991)}}.

%%CITATION = 0806.1519;%%
\bibitem{Benna:2008zy}
M.~Benna, I.~Klebanov, T.~Klose and M.~Smedback,
\textit{``{Superconformal Chern-Simons theories and $AdS_4/CFT_3$
  correspondence}''},
\textsf{\doiref{10.1088/1126-6708/2008/09/072}{JHEP~0809,~072~(2008)}},
\texttt{\arxivref{0806.1519}{arxiv:0806.1519}}.

%%CITATION = HEP-PH 9309335;%%
\bibitem{Seiberg:1993vc}
N.~Seiberg,
\textit{``Naturalness versus supersymmetric nonrenormalization theorems''},
\textsf{\doiref{10.1016/0370-2693(93)91541-T}{Phys.~Lett.~B318,~469~(1993)}},
\texttt{\arxivref{hep-ph/9309335}{hep-ph/9309335}}.

%%CITATION = PHLTA,B281,72;%%
\bibitem{Gates:1991qn}
S.~J.~Gates and H.~Nishino,
\textit{``Remarks on the $\CN=2$ supersymmetric Chern-Simons theories''},
\textsf{\doiref{10.1016/0370-2693(92)90277-B}{Phys.~Lett.~B281,~72~(1992)}}.

%%CITATION = HEP-TH/9401053;%%
\bibitem{Kapustin:1994mt}
A.~N.~Kapustin and P.~I.~Pronin,
\textit{``{Nonrenormalization theorem for gauge coupling in (2+1)-
  dimensions}''},
\textsf{\doiref{10.1142/S0217732394001787}{Mod.~Phys.~Lett.~A9,~1925~(1994)}},
\texttt{\arxivref{hep-th/9401053}{hep-th/9401053}}.

%%CITATION = HEP-TH/9711191;%%
\bibitem{DelCima:1997pb}
O.~M.~Del~Cima, D.~H.~T.~Franco, J.~A.~Helayel-Neto and O.~Piguet,
\textit{``{On the non-renormalization properties of gauge theories with
  Chern-Simons terms}''},
\textsf{\doiref{10.1088/1126-6708/1998/02/002}{JHEP~9802,~002~(1998)}},
\texttt{\arxivref{hep-th/9711191}{hep-th/9711191}}.

%%CITATION = NUPHA,B334,103;%%
\bibitem{AlvarezGaume:1989wk}
L.~Alvarez-Gaume, J.~M.~F.~Labastida and A.~V.~Ramallo,
\textit{``A note on perturbative Chern-Simons theory''},
\textsf{\doiref{10.1016/0550-3213(90)90658-Z}{Nucl.~Phys.~B334,~103~(1990)}}.

%%CITATION = HEP-TH/9209005;%%
\bibitem{Chen:1992ee}
W.~Chen, G.~W.~Semenoff and Y.-S.~Wu,
\textit{``{Two loop analysis of nonAbelian Chern-Simons theory}''},
\textsf{\doiref{10.1103/PhysRevD.46.5521}{Phys.~Rev.~D46,~5521~(1992)}},
\texttt{\arxivref{hep-th/9209005}{hep-th/9209005}}.

%%CITATION = HEP-TH/9506170;%%
\bibitem{Kao:1995gf}
H.-C.~Kao, K.-M.~Lee and T.~Lee,
\textit{``{The Chern-Simons coefficient in supersymmetric Yang-Mills
  Chern-Simons theories}''},
\textsf{\doiref{10.1016/0370-2693(96)00119-0}{Phys.~Lett.~B373,~94~(1996)}},
\texttt{\arxivref{hep-th/9506170}{hep-th/9506170}}.

%%CITATION = 0806.4900;%%
\bibitem{Bedford:2008hn}
J.~Bedford and D.~Berman,
\textit{``{A note on quantum aspects of multiple membranes}''},
\textsf{\doiref{10.1016/j.physletb.2008.08.021}{Phys.~Lett.~B668,~67~(2008)}},
\texttt{\arxivref{0806.4900}{arxiv:0806.4900}}.

%%CITATION = HEP-TH/0108200;%%
\bibitem{Gates:1983nr}
S.~J.~Gates, M.~T.~Grisaru, M.~Rocek and W.~Siegel,
\textit{``{Superspace, or one thousand and one lessons in supersymmetry}''},
\textsf{Front.~Phys.~58,~1~(1983)},
\texttt{\arxivref{hep-th/0108200}{hep-th/0108200}}.

\bibitem{Buchbinder:1998qv}
I.~L.~Buchbinder and S.~M.~Kuzenko,
\textit{``Ideas and methods of supersymmetry and supergravity: Or a walk
  through superspace''},
Bristol, UK: IOP (1998) 656 p.

%%CITATION = NUPHA,B382,561;%%
\bibitem{Avdeev:1991za}
L.~V.~Avdeev, G.~V.~Grigorev and D.~I.~Kazakov,
\textit{``{Renormalizations in abelian Chern-Simons field theories with
  matter}''},
\textsf{\doiref{10.1016/0550-3213(92)90659-Y}{Nucl.~Phys.~B382,~561~(1992)}}.

%%CITATION = PHLTA,B84,193;%%
\bibitem{Siegel:1979wq}
W.~Siegel,
\textit{``Supersymmetric dimensional regularization via dimensional
  reduction''},
\textsf{\doiref{10.1016/0370-2693(79)90282-X}{Phys.~Lett.~B84,~193~(1979)}}.

%%CITATION = NUPHA,B159,429;%%
\bibitem{Grisaru:1979wc}
M.~T.~Grisaru, W.~Siegel and M.~Rocek,
\textit{``Improved methods for supergraphs''},
\textsf{\doiref{10.1016/0550-3213(79)90344-4}{Nucl.~Phys.~B159,~429~(1979)}}.

%%CITATION = HEP-TH/0609149;%%
\bibitem{Ananth:2006ac}
S.~Ananth, S.~Kovacs and H.~Shimada,
\textit{``{Proof of all-order finiteness for planar beta-deformed
  Yang-Mills}''},
\textsf{\doiref{10.1088/1126-6708/2007/01/046}{JHEP~0701,~046~(2007)}},
\texttt{\arxivref{hep-th/0609149}{hep-th/0609149}}.

%%CITATION = HEP-TH/0112258;%%
\bibitem{Witten:2001ua}
E.~Witten,
\textit{``Multi-trace operators, boundary conditions, and AdS/CFT
  correspondence''},
\texttt{\arxivref{hep-th/0112258}{hep-th/0112258}}.

\end{thebibliography}
\end{document}